\documentclass[sigconf]{acmart}

\usepackage{hyperref} 
\usepackage{hyperxmp} 
\usepackage{microtype}
\usepackage{fancyvrb} 
\usepackage{color}
\usepackage{enumitem}
\setlist[itemize]{leftmargin=11pt}

\AtBeginDocument{%
  \providecommand\BibTeX{{%
    \normalfont B\kern-0.5em{\scshape i\kern-0.25em b}\kern-0.8em\TeX}}}

\definecolor{lightgrey}{rgb}{0.8,0.8,0.8}
\definecolor{white}{rgb}{1.0,1.0,1.0}

\newcommand{\orbit}{{\texttt{ORBIT}}}

\newcommand{\cmark}{\ding{51}}
\newcommand{\xmark}{\ding{55}}
\newcommand{\orbitm}{\orbit{}\textsubscript{Ext}}
\newcommand{\orbita}{\orbit{}\textsubscript{Gen}}

\definecolor{hlrow}{rgb}{0.68,0.85,0.90}        

\newcommand{\hl}[1]{\cellcolor{hlrow}#1}

\usepackage{booktabs}
\usepackage{makecell}
\usepackage{tabularx}

\usepackage{array}
\usepackage{amsmath}
\usepackage{microtype}
\usepackage{graphicx}
\usepackage{float}
\usepackage{xspace}
\usepackage{enumitem}
\setlistdepth{9}

\usepackage{lipsum}
\usepackage{todonotes}
\usepackage{subcaption}
\usepackage{fancyhdr}
\usepackage{comment}
\usepackage{tcolorbox}
\usepackage{listings}
\usepackage{xcolor}
\usepackage{caption}
\usepackage{booktabs}
\usepackage{multirow}
\usepackage{rotating}
\usepackage{geometry}
\usepackage{array}
\usepackage{pifont}
\usepackage{colortbl}

\usepackage[ruled, vlined, linesnumbered]{algorithm2e}
\SetAlFnt{\small}
\SetAlgoSkip{}
\SetAlCapFnt{\small}
\SetAlCapNameFnt{\small}
\SetNlSty{textnormal}{}{:}
\SetKwRepeat{Repeat}{repeat}{until}
\newcommand{\CompileRepair}{\textsc{Compile{\normalfont\&}Repair}}
\newcommand{\VerifyRepair}{\textsc{Verify{\normalfont\&}Repair}}
\newcommand{\Repair}{\textsc{Repair}}

\definecolor{rulecolor}{RGB}{255,0,0}
\definecolor{commentcolor}{RGB}{128,128,128} 
\definecolor{placeholdercolor}{RGB}{0,128, 0} 

\tcbset{
  prompt box/.style={
    colframe=lightgray,
    colback=white,
    boxrule=0.5pt,
    arc=0pt,
    outer arc=0pt,
    fontupper=\sffamily, 
    before upper={\sffamily}, 
    left=4pt,    
    right=4pt,   
    top=4pt,     
    bottom=4pt,  
    }
}

\definecolor{codebg}{rgb}{0.95,0.95,0.95}
\definecolor{keyword}{rgb}{0.0,0.0,0.5}
\definecolor{string}{rgb}{0.5,0.0,0.0}
\definecolor{comment}{rgb}{0.25,0.5,0.25}
\definecolor{function}{rgb}{0.0,0.0,0.6}
\definecolor{number}{rgb}{0.0,0.0,0.6}

\definecolor{mygreen}{rgb}{0,0.6,0}
\definecolor{mygray}{rgb}{0.5,0.5,0.5}
\definecolor{mymauve}{rgb}{0.58,0,0.82}

\lstdefinelanguage{Rust}{
  keywords={fn, let, mut, pub, const, as, use, self, impl, trait, match, enum, struct, ref, if, else, loop, while, for, in, return, break, continue},
  keywordstyle=\color{blue}\bfseries,
  ndkeywords={i32, u32, usize, Result, Option, Some, None, Ok, Err, String, Vec},
  ndkeywordstyle=\color{mymauve}\bfseries,
  identifierstyle=\color{black},
  sensitive=true,
  comment=[l]{//},
  morecomment=[s]{/*}{*/},
  commentstyle=\color{mygreen}\ttfamily,
  stringstyle=\color{mymauve}\ttfamily,
  morestring=[b]',
  morestring=[b]",
}

\lstdefinelanguage{C}{
  language=C,
  morekeywords={uint8_t, uint16_t, uint32_t, int8_t, int16_t, int32_t, size_t},
  keywordstyle=\color{blue}\bfseries,
  ndkeywords={NULL},
  ndkeywordstyle=\color{mymauve}\bfseries,
  identifierstyle=\color{black},
  sensitive=true,
  comment=[l]{//},
  morecomment=[s]{/*}{*/},
  commentstyle=\color{mygreen}\ttfamily,
  stringstyle=\color{mymauve}\ttfamily,
  morestring=[b]',
  morestring=[b]",
}

\microtypecontext{spacing=nonfrench}

\hypersetup{
  bookmarks=false, 
  colorlinks=true,%
  citecolor=black,%
  filecolor=black,%
  linkcolor=black,%
  urlcolor=black,%
  pagebackref=true,%
}

\setcopyright{none}
\acmConference{}{}{}
\acmDOI{}
\acmISBN{}
\renewcommand\footnotetextcopyrightpermission[1]{}

\begin{document}

\title{ORBIT: Guided Agentic Orchestration for Autonomous\\C-to-Rust Transpilation}



\author{Muhammad Farrukh}
    \affiliation{%
        \institution{Stony Brook University}
        \city{New York}
        \country{USA}
    }
    \email{mufarrukh@cs.stonybrook.edu}

\author{Baris Coskun}
    \authornote{This work does not relate to Baris Coskun's position at Amazon.}
    \affiliation{%
        \institution{Amazon Web Services}
        \city{New York}
        \country{USA}
    }
        \email{barisco@amazon.com}

\author{Tapti Palit}
    \affiliation{%
        \institution{University of California, Davis}
        \city{Davis}
        \country{USA}
    }
    \email{tpalit@ucdavis.edu}

\author{Michalis Polychronakis}
    \affiliation{%
        \institution{Stony Brook University}
        \city{New York}
        \country{USA}
    }
    \email{mikepo@cs.stonybrook.edu}





\begin{abstract} Large-scale migration of legacy C code to Rust offers a
promising path toward improving memory safety, but LLM-based C-to-Rust
translation remains challenging due to limited context windows and
hallucinations. Prior approaches are evaluated primarily on small programs or
datasets skewed toward small codebases, providing limited insight into
scalability on real-world systems. They also rely on static context
construction, which breaks down in the presence of complex cross-module
dependencies and often requires manual intervention. Recent coding agents offer
a promising alternative through dynamic codebase navigation and context
curation. When used out of the box, however, they frequently produce incomplete
translations that appear superficially correct.

We present ORBIT, an autonomous agentic framework for project-level C-to-Rust
translation that combines dynamic context collection with dependency-guided
orchestration and iterative verification. ORBIT constructs a dependency-aware
translation graph, generates Rust interfaces, maps C functions to Rust
counterparts, and coordinates multiple specialized agents. We evaluate ORBIT on
24 programs from CRUST-Bench, with 91.7\% of the programs exceeding 1,000 lines of
code. ORBIT achieves 100\% compilation success and 91.7\% test success in both
expert-interface and automatically generated-interface settings, substantially
outperforming C2Rust and CRUST-Bench, while reducing unsafe Rust code blocks to nearly
zero. We further evaluate ORBIT on challenging cases from the DARPA TRACTOR
benchmark, where it achieves competitive performance relative to participating
systems.  \end{abstract}
\maketitle
\pagestyle{plain}

\section{Introduction}
\label{sec:introduction}

The widespread use of legacy C code continues to expose critical
systems to memory safety vulnerabilities. Microsoft has reported that
70\% of CVEs originate from memory corruption~\cite{MicrosoftMSRC2019}, and
decades of effort to integrate memory safety mechanisms into C/C++
have failed to achieve broad adoption due to performance and
compatibility constraints. In contrast, Rust is a modern systems language that enforces memory safety at compile time through its ownership and borrowing model, with a growing effort to make it secure and mature from industry and academia~\cite{AWS_Kani_2022,hassnain2024counterexamples, hassnain2026cargosherlocksmtbasedchecker}. This has motivated significant
interest in migrating critical software systems from C to Rust. Recent
initiatives such as DARPA's TRACTOR program~\cite{DARPA_TRACTOR} and the
Great Refactor~\cite{TheGreatRefactor2024} reflect a systematic community effort
to address this challenge at scale. Automated C-to-Rust translation
offers a promising path toward this goal, enabling large-scale
modernization without requiring complete manual rewriting.

Earlier automated approaches adopted predefined rule-based
transpilation techniques~\cite{immunant2022c2rust, crown,
hong2023concrat, hong2024tag, safer-rust} that scale to large
codebases without incurring model inference costs. However, these
methods produce unsafe and unidiomatic Rust code that is difficult to
maintain and forfeits the safety guarantees that motivate 
migration in the first place. Recent work has therefore shifted
towards leveraging large language models (LLMs) as the primary
translation engine, owing to their ability to generate idiomatic and
safe code. Despite their impressive abilities,
LLM-based approaches face two principal
challenges: (i)~\textit{semantic correctness}, which recent studies
address through various verification and repair
techniques~\cite{yang2024unitrans, eniser2024flourine,
nitin2024spectra, yang2024vert, bai2025rustassure}; and
(ii)~\textit{limited context window}, which restricts the amount of
code that can be translated in a single inference call. A substantial
body of work~\cite{nitin2025c2saferrust, shetty2024syzygy,
shiraishi2024context, zhang2024scalable} has addressed the limitation of context, yet two significant gaps remain.

\textbf{Limited evaluation scale.} Although recent
tools~\cite{shetty2024syzygy, shiraishi2024smartc2rust, evoc2rust, cai2025rustmap} claim to handle larger C programs
($>$500~LoC), their evaluation benchmarks consist predominantly of
small programs, with only a small fraction exceeding 1,000~LoC.
Table~\ref{tab:dataset-comparison} provides a comprehensive comparison
across tools, showing that most prior works operate on datasets with
median program sizes well below 1K~LoC. Some tools do not report
dataset sizes explicitly, therefore we only present those for which we can provide best-effort approximations.
We exclude LLM-based approaches that refine C2Rust
output~\cite{nitin2025c2saferrust, gao2025pr2},
because the resulting translation still contains a large amount of unsafe
code blocks compared to approaches that directly start with LLM output, and
Tymcrat~\cite{hong2025type}, which targets large repositories but
performs type migration on function subsets rather than complete
translation. Such evaluations fail to capture the structural
complexity and cross-module dependencies of real-world
codebases.

\textbf{Complex static analysis for program decomposition.} 
To address the context window limitation, several approaches employ static
analysis to curate the dependencies of translation units,
(e.g., helper functions, type definitions, and macros)
and perform incremental translation in dependency order.
However, Li et al.~\cite{li2024userstudy} report
that VERT~\cite{yang2024vert} and Flourine~\cite{eniser2024flourine}
required substantial manual effort to produce an initial translation
on a new dataset. Similarly, several
works~\cite{shetty2024syzygy, cai2025rustmap,
10.1145/3735544.3735582} require human intervention to resolve
complex dependencies or struct types. Beyond that, the use of complex techniques to aid the LLM during translation, such as lifetime dependency tracking, pointer aliasing, and ownership analysis, struggle due to the inherent complexities of static analysis. These limitations reflect a
broader challenge: context construction
is largely static and predetermined prior to translation, leaving
systems unable to adapt when the initially curated context is
incomplete or misaligned with the translation task. Effective
C-to-Rust translation therefore requires not only context-aware
generation, but also the ability to dynamically acquire, validate,
and refine context during execution.

Recent advances in coding agents provide a natural abstraction for
addressing both gaps. Modern agentic systems such as GitHub Copilot,
Codex, and Claude Code have become deeply integrated into the software
development lifecycle~\cite{robbes2026agentic, li2025rise, babar2025open,
BusinessInsiderUberAI2026}.
By enabling tool interaction and
feedback-driven refinement, agentic systems can dynamically gather
missing context, resolve dependencies, and adapt translation
strategies based on intermediate failures. However, in our
preliminary experiments we observed that vanilla coding agents used in
a single-shot setting frequently produce incomplete
translations while misleadingly reporting success---as shown in Section~\ref{sec:ablation}, the base agent achieves build and test success while covering only 64.4\% of functions and 3.0\% of test cases. This behavior is particularly pronounced in open-source
models. Furthermore, even when agents do produce translations,
Khatry et al.~\cite{khatry2025crust} report that agents perform a disproportionately
large number of codebase navigation steps, suggesting that careful
orchestration is required to unlock their potential as autonomous
translation systems.

In this paper, we present \orbit{}, an agentic framework for
C-to-Rust translation that addresses these limitations. 
Unlike prior approaches
that rely on complex static analysis for context construction, \orbit{}
employs lightweight call-site analysis for ordering and delegates
context curation to the agent itself, allowing the system to
dynamically adapt to intermediate failures. \orbit{} supports both
expert-provided Rust interfaces and interfaces generated automatically
via Agentic Iterative Scaffolding, making it applicable without
manual input.

 
 
 
 
 
 
 
 
 

We evaluate \orbit{} on a benchmark of 24 programs derived from
CRUST-Bench~\cite{khatry2025crust}, with 91.7\% of them exceeding 1,000
LoC (Table~\ref{tab:dataset-comparison}), using two state-of-the-art
agents: Opencode with Qwen3-Coder-480B and Codex with
GPT-5.2-Codex. When run with Qwen, which is less capable than GPT-Codex, \orbit{} achieves 100\% compilation success and
91.7\% test success on all 24 CRUST-Bench programs in both interface
configurations, substantially outperforming C2Rust (62.5\%
compilation, 20.8\% test success) and CRUST-Bench (45.8\%
compilation, 25.0\% test success). On the safety dimension, \orbit{}
reduces the average number of unsafe lines of code
to 0.06\% and 0.11\% under the two
interface configurations, compared to 69.6\% for C2Rust. We further
evaluate \orbit{} on the DARPA TRACTOR benchmark, achieving a pass
rate of $\sim$70\% (9/13) on the hardest programs in the battery,
placing it competitively within the range of six performer
systems.

In summary, our main contributions are as follows:
\begin{itemize}
    \item We present \orbit{}, a fully autonomous agentic framework
    for translating large-scale C projects into functional and
    memory-safe Rust, requiring no manual intervention in the
    translation process.

    \item We present the first rigorous evaluation of an
    LLM-based C-to-Rust translation tool on a benchmark where over
    90\% of the programs exceed 1,000~LoC, demonstrating scalability to
    real-world codebases.

    \item We systematically demonstrate that vanilla coding agents yield incomplete translations that appear correct under standard build and test metrics, exposing a critical gap between superficial correctness and true functional completeness.
\end{itemize}

\begin{table}[t]
    \centering
    \scriptsize
    \setlength{\tabcolsep}{3pt}
    \setlength{\belowbottomsep}{0pt}
    \caption{Comparison of datasets used in prior C-to-Rust translation works.}
    \label{tab:dataset-comparison}
    \renewcommand{\arraystretch}{1.0}
\begin{tabular}{
p{1.6cm}
p{2.0cm}
>{\centering\arraybackslash}m{0.6cm}
>{\centering\arraybackslash}m{0.7cm}
>{\centering\arraybackslash}m{0.8cm}
>{\centering\arraybackslash}m{0.8cm}
}
    \toprule
    \textbf{Tool} & \textbf{Dataset} & \textbf{\#Prog} & \textbf{\%$>$1K}
      & \textbf{Med} & \textbf{Mean} \\
    \midrule
    RustMap~\cite{cai2025rustmap}      & Rosetta Code, Bzip2         & 126 & $\sim$1\%  & $\sim$80    & $\sim$145  \\
    Syzygy~\cite{shetty2024syzygy}     & Zopfli, URL parser          & 2   & 50\%       & 2700        & 2700       \\
    EvoC2Rust~\cite{evoc2rust}         & C2R-Bench, Vivo-Bench       & 25  & $\sim$8\%  & $\sim$400   & $\sim$600  \\
    RustAssure~\cite{bai2025rustassure}& 5 C Libs                    & 5   & 20\%       & 405         & $\sim$900  \\
    SmartC2Rust~\cite{shiraishi2024smartc2rust} & GitHub, Prior Studies Data & 21 & $\sim$24\% & 502 & $\sim$1000 \\
    VERT~\cite{yang2024vert}           & TransCoder-IR, Prior Studies & 534 & 0\%       & $\sim$100   & $\sim$120  \\
    \midrule
    \textbf{ORBIT} & \textbf{CRUST-Bench} & \textbf{24} & \textbf{91.7\%} & \textbf{1,354} & \textbf{1,603} \\
    \bottomrule
\end{tabular}
\end{table}
\section{ORBIT Overview}
\label{sec:methodology}



\begin{figure*}[t]
    \centering
    \includegraphics[width=0.85\textwidth]{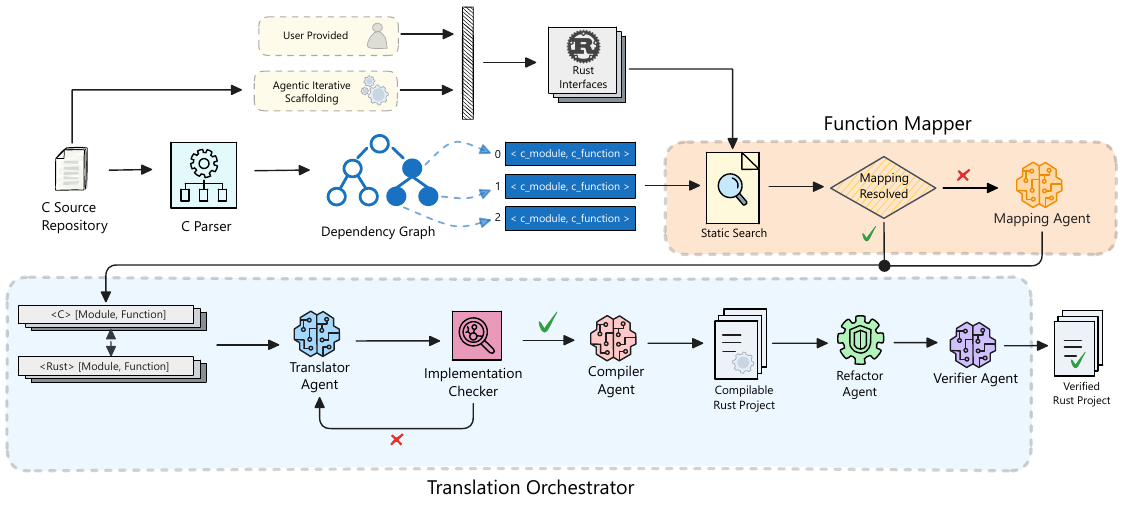}
    \caption{Overview of the \orbit{}.}
    \label{fig:orbit-overview}
\end{figure*}

In this section, we describe the architectural details of \orbit{}. The primary
design goal of \orbit{} is to leverage the capabilities of coding agents to
abstract away the engineering complexities of automated translation, while
decomposing the translation task into a structured, multi-stage format. \orbit{}
consists of four main components as illustrated in Figure~\ref{fig:orbit-overview}: (i)~\emph{Sorted Dependency Graph
Generator}, which parses C source code into a dependency graph and produces a
topologically sorted ordering at both the module and function levels;
(ii)~\emph{Agentic Iterative Scaffolding}, which creates a memory-safe Rust project
skeleton corresponding to a given C repository containing translated data types,
interfaces, and test suites; (iii)~\emph{Function Mapper}, which takes the
sorted C modules and functions alongside Rust interfaces to establish a
correspondence between them, used subsequently during translation;
and (iv)~\emph{Translation Orchestrator}, which coordinates multiple specialized agents
and performs deterministic verification on agent-generated outputs.

\subsection{Sorted Dependency Graph Generator}

Given a source C repository, \orbit{} employs a parser built on Tree-sitter~\cite{tree-sitter}
to generate an Abstract Syntax Tree (AST) for each \texttt{.c} and \texttt{.h} file
in the codebase. Rather than tracking all source code constructs such as
variables, structs, and type definitions, the parser focuses exclusively on two
primary entity types: (1)~\textit{Function Nodes}, encompassing standard
function signatures, macro-wrapped names, and static inline functions;
and (2)~\textit{Include Nodes}, which capture file-level dependencies. From these
entities, \orbit{} constructs a dual-layered dependency graph comprising two
sub-graphs: (a)~an \textit{Intra-Module Graph}, which captures local
dependencies between functions within the same source or header file (module);
and (b)~an \textit{Inter-Module Graph}, which tracks high-level dependencies
between \textit{Module Groups}, defined as logical groupings of related
\texttt{.c} and \texttt{.h} files. To ensure that foundational entities are processed before those that depend on them (a requirement for both correct translation order and successful compilation) \orbit{} applies Kahn's algorithm~\cite{kahn1962topological} in
two stages:
\begin{itemize}
    \item \emph{Intra-Module Sort:} Functions within a single file are ordered
    such that callees precede their callers wherever the dependency structure
    permits.
    \item \emph{Inter-Module Sort:} Modules are ordered according to the
    inter-module dependency graph, ensuring that base libraries and utility
    headers are processed prior to the higher-level logic that depends on them.
\end{itemize}

This two-stage sorting establishes a dependency-aware translation order and lays
the groundwork for constructing C-to-Rust function mappings in subsequent
stages. Since C projects frequently contain circular dependencies, \orbit{} detects such
cycles and groups the involved entities into a single translation unit,
preserving correctness while maintaining translation progress.

\subsection{Agentic Iterative Scaffolding}
This component is responsible for generating a type-checked Rust project
skeleton, which serves as a precursor to the main translation loop. Rather than
relying on complex static analysis to discover translation units, such as global
variables and type definitions, \orbit{} employs an off-the-shelf coding agent to
perform the scaffolding directly. Recent advancements in coding agents enable
them to automatically curate the necessary context based on the task at hand.
However, they remain prone to failure when presented with a single large
monolithic task. Therefore, \orbit{} relies on a multi-stage iterative process
to guide the agent through the scaffolding incrementally.
Algorithm \ref{alg:interface} outlines the key steps in this process:

\emph{Project Skeleton Initialization (Lines 1--3):} \orbit{} runs \texttt{cargo new} and creates Rust modules mirroring each C source file; header files with non-declaration content are preserved with a \texttt{\_h} suffix.
\emph{Type Translation (Lines 4--6):} C type definitions (\texttt{struct}, \texttt{union}, \texttt{enum}, macros, primitives) are mapped into the corresponding Rust modules.
\emph{Function Signature Resolution (Lines 7--10):} Rust function signatures are generated for each C function, with bodies stubbed as \texttt{unimplemented!()}, producing a structurally complete skeleton for downstream mapping.
\emph{Safety Refactoring (Lines 11--15):} Raw pointer signatures and unsafe constructs are iteratively refactored towards safe Rust abstractions.
\emph{Test Suite Translation (Lines 16--18):} The C test suite is translated into Rust \texttt{\#[test]} functions and integrated into the \texttt{cargo test} harness.
\emph{Verification (Lines 19--20):} A final \VerifyRepair{} pass resolves residual structural errors, yielding a type-checked skeleton $\mathcal{R}$.

\begin{algorithm}[t]
\scriptsize 
\caption{Agentic Iterative Scaffolding}
\label{alg:interface}
\KwIn{$\mathcal{C}$: C repository; $\mathcal{A}$: coding agent; $N$: max repair
attempts; $K$: max refactor attempts; $T_{out}$: stage timeout}
\KwOut{$\mathcal{R}$: compilable Rust skeleton} $\mathcal{R} \leftarrow
\emptyset$\; $\mathcal{R} \leftarrow \textsc{CargoNew}(\mathcal{C},\
\mathcal{A},\ T_{out})$\; $\mathcal{R} \leftarrow \CompileRepair(\mathcal{R},\
\mathcal{A},\ N,\ T_{out})$\; \While{$\mathcal{A}$ \textnormal{identifies} $t
\in \mathcal{C}_{types}\ \land\ \neg\textsc{TimedOut}(T_{out})$}{ $\mathcal{R}
\leftarrow \mathcal{R}\ \cup\ \{\textsc{MapType}(t)\}$\; } $\mathcal{R}
\leftarrow \CompileRepair(\mathcal{R},\ \mathcal{A},\ N,\ T_{out})$\;
\While{$\mathcal{A}$ \textnormal{identifies} $f \in \mathcal{C}_{fns}\ \land\
\neg\textsc{TimedOut}(T_{out})$}{ $sig_f \leftarrow \textsc{MapSignature}(f)$\;
$\mathcal{R} \leftarrow \mathcal{R}\ \cup\ \{sig_f \Rightarrow
\texttt{unimplemented!()}\}$\; } $\mathcal{R} \leftarrow
\CompileRepair(\mathcal{R},\ \mathcal{A},\ N,\ T_{out})$\; $k \leftarrow 0$\;
\While{$\neg\textsc{IsSafe}(\mathcal{R})\ \land\ k < K$}{ $\mathcal{R}
\leftarrow \textsc{Refactor}(\mathcal{R},\ \mathcal{A},\ T_{out})$\; $k
\leftarrow k + 1$\; } $\mathcal{R} \leftarrow \CompileRepair(\mathcal{R},\
\mathcal{A},\ N,\ T_{out})$\; \While{$\mathcal{A}$ \textnormal{identifies} $\tau
\in \mathcal{C}_{tests}\ \land\ \neg\textsc{TimedOut}(T_{out})$}{ $\mathcal{R}
\leftarrow \mathcal{R}\ \cup\ \{\texttt{\#[test]}\ \textsc{MapTest}(\tau)\}$\; }
$\mathcal{R} \leftarrow \CompileRepair(\mathcal{R},\ \mathcal{A},\ N,\
T_{out})$\; $\mathcal{R} \leftarrow \VerifyRepair(\mathcal{R},\ \mathcal{A},\
N,\ T_{out})$\; \Return $\mathcal{R}$\;
\SetKwProg{Proc}{Function}{}{} \Proc{$\CompileRepair(\mathcal{R},\ \mathcal{A},\
N,\ T_{out})$}{ $k \leftarrow 0$\;
\While{$\neg\textsc{IsCompilable}(\mathcal{R})\ \land\ k < N$}{ $\mathcal{R}
\leftarrow \Repair(\mathcal{R},\ \mathcal{A},\ \textsc{GetErrors}(\mathcal{R}),\
T_{out})$\; $k \leftarrow k + 1$\; } \Return $\mathcal{R}$\; }
\end{algorithm}

\subsection{Function Mapper}

Establishing a precise correspondence between C functions and their Rust
counterparts is non-trivial due to the fundamental idiomatic differences between
the two languages. The Translation Orchestrator requires an exact, unambiguous
mapping from every C function to its corresponding Rust stub in the skeleton, as
this mapping instructs the translation agent precisely which stub to implement
at each step. Without it, the agent has no well-defined target and risks
missing function implementations.

\begin{figure}[t]
    \centering
    \includegraphics[width=0.50\textwidth]{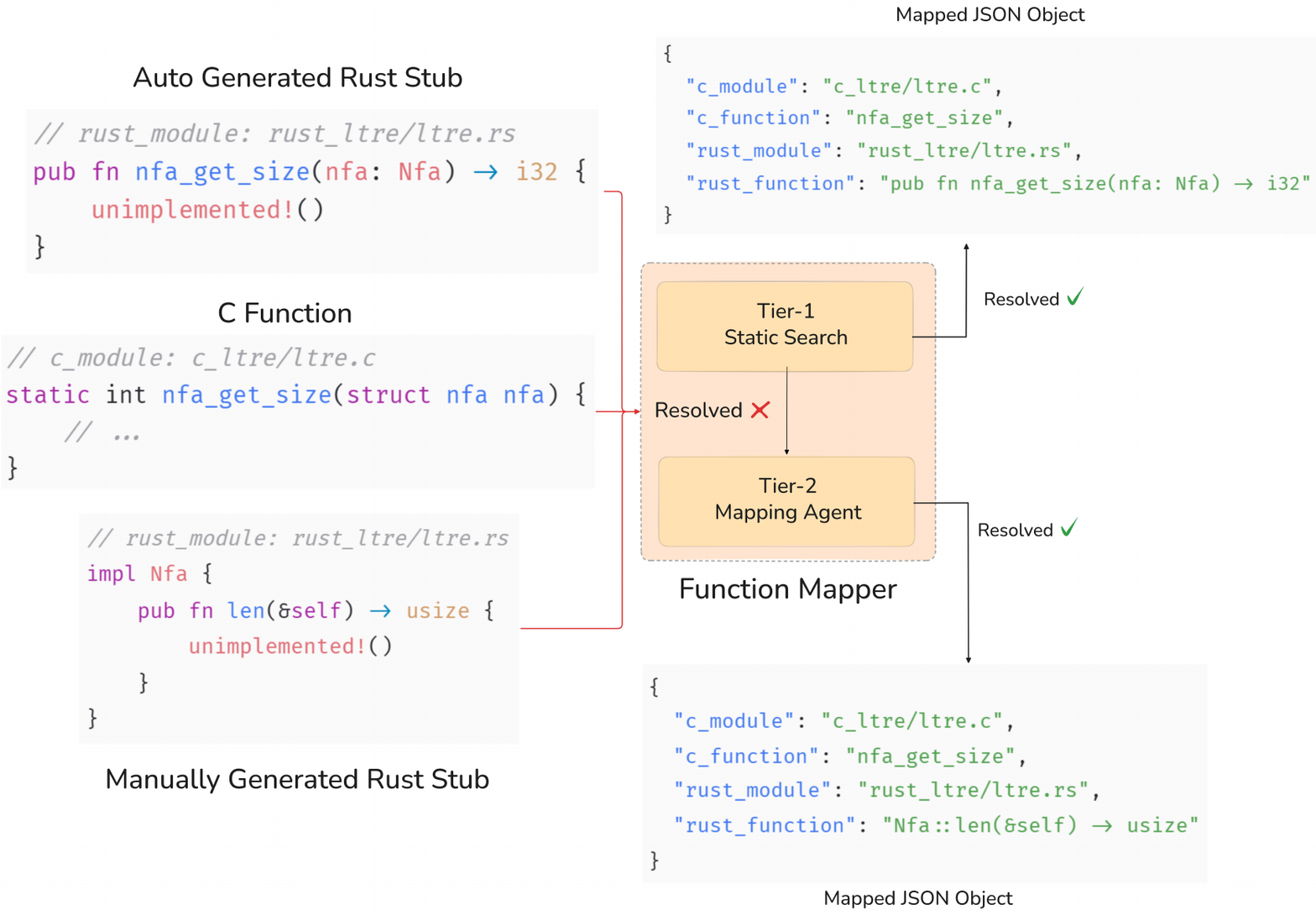}
    \caption{Two-tier function mapping strategy in \orbit{}.
    }
    \label{fig:mapper}
\end{figure}

This challenge is most pronounced when using manually curated Rust skeletons.
Figure~\ref{fig:mapper} illustrates an example using both auto-generated and
manually curated Rust stubs derived from a C function in the \texttt{LTRE}
program from CRUST-Bench~\cite{khatry2025crust}.
In the manually curated interface, the C function
\texttt{nfa\_get\_size} (which operates on a struct) is lifted into an
associated method \texttt{Nfa::len}, fundamentally reorganising the function
structure. In such cases, string-matching heuristics fail to recover the correct
correspondence or lead to false positives. As a secondary concern, surface-level
naming inconsistencies compound the problem further. For instance, in the same
program, \texttt{bitset\_get} is preserved in the auto-generated interface but
is renamed to \texttt{bitset\_test} in the manual one.

To handle both cases robustly without requiring any configuration changes,
\orbit{} employs a two-tier mapping strategy: a fast, deterministic static
search as the primary tier, with a Mapping Agent fallback invoked only when all
static matching strategies are exhausted. As illustrated in
Figure~\ref{fig:mapper}, when the auto-generated skeleton is used, the static
search resolves \texttt{nfa\_get\_size} directly via exact-match lookup. When
the manually curated skeleton is used, the same function cannot be resolved
statically due to the free-function-to-method transformation, and the Mapping
Agent is invoked to reason over the semantic correspondence between
\texttt{nfa\_get\_size} and \texttt{Nfa::len}. This design ensures that \orbit{}
is agnostic to the origin of the Rust skeleton, supporting both auto-generated
and manually curated interfaces within a unified pipeline.

\subsubsection{Tier 1: Static Search}
For each (\texttt{c\_module}, \texttt{c\_function})
pair drawn from the sorted topology,
\orbit{} first attempts a grammar-based lookup using \texttt{ripgrep} against
the Rust project. The search applies a strict matching hierarchy: it first
attempts an exact literal match of the C function name against Rust \texttt{fn}
declarations, followed by a normalised match in which both names are lowercased
and stripped of underscores and hyphens prior to comparison. Looser forms of
matching such as substring or prefix matching are deliberately excluded to
prevent false-positive mappings.

\subsubsection{Tier 2: Mapping Agent Fallback}
If static search yields no result, the Mapping Agent is invoked with a
structured prompt containing the target C function signature and its source
module, alongside with full access to C and Rust projects. The agent reasons
over semantic correspondence, parameter count, parameter types, and return types
to identify structurally transformed counterparts, returning a structured JSON
response identifying the matched \texttt{rust\_module} and
\texttt{rust\_function}. In cases where no Rust equivalent exists, such as C
deallocator functions that have no counterpart in Rust due to its built-in
ownership and deallocation model, the Mapping Agent assigns \texttt{null} to the fields.

Since LLMs are inherently prone to hallucination, every mapping resolved by the
Mapping Agent passes through a validation step before being committed. The
validator confirms that both the \texttt{c\_module} and \texttt{rust\_module}
paths actually exist and that the mapped \texttt{rust\_function} is present
within the specified Rust file via a targeted static search. This prevents
hallucinated mappings from propagating into the translation phase. If validation
fails, the Mapping Agent is reinvoked for up to a specified number of attempts.
\vspace{-5ex}
\subsection{Translation Orchestrator}

Since functions are translated incrementally in dependency order, each translation step must be verified before proceeding, partial failures must be recoverable, and the overall process must remain resilient to the non-deterministic behavior of coding agents. The decomposition of translation into distinct, specialized agents is motivated by prior work demonstrating that role-specialized multi-agent systems consistently outperform monolithic
single-agent approaches on complex software engineering
tasks~\cite{hong2024metagpt, huang2023agentcoder, qian2024chatdev}. The Translation Orchestrator addresses these concerns by imposing a structured, iterative execution loop over the function mappings produced by the Function Mapper, treating each (C function, Rust stub) pair as an independent translation unit.

For each translation unit, the Translation Agent is invoked to generate the Rust
implementation of the target C function. The agent is provided with a
translation prompt, an \texttt{AGENTS.md} file, and access to both the C and
Rust projects. We observed that coding agents tend to perform actions beyond the
scope of the prompt, such as prematurely running tests before the implementation
is complete. To mitigate this, we analyzed agent behaviour on a set of sample
prompts and introduced explicit constraints into the prompt to suppress such
over-eager actions.

Once the agent completes the implementation, an Implementation Checker is
invoked to verify that the target Rust function has been genuinely implemented.
This checker uses static analysis to confirm that the function body contains
neither an \texttt{unimplemented!()} macro nor TODO comments. The rationale for
this explicit verification step stems from an observed failure mode: when used
with certain open-source models, agents would simply remove the
\texttt{unimplemented!()} macro and replace the function body with TODO
comments, satisfying the surface form of the prompt without producing a valid
implementation. If the Implementation Checker fails, the Translation Agent is
reinvoked for up to a specified number of attempts.

Once a function is successfully implemented, the Translation Orchestrator
invokes the Compiler Agent to ensure that the Rust project compiles successfully
with the new implementation. If compilation errors are detected, the Compiler
Agent attempts to repair the code iteratively until the project compiles or the
maximum number of repair attempts is reached. Upon successful compilation, the
Translation Orchestrator records the translation unit as completed and advances
to the next.

After all translation units have been processed, a Refactoring Agent is invoked
to eliminate unsafe operations introduced during translation. When this agent
returns, the Rust project is passed to the Verifier Agent, which runs the test
suite and repairs any failing test cases within the allotted number of attempts.
The result is a complete, verified Rust translation of the source C codebase.

A key design principle of \orbit{} is that all agents operating under the
Translation Orchestrator are paired with deterministic verification steps.
Rather than accepting an agent's output at face value, \orbit{} independently
confirms the result using static analysis or toolchain checks. For instance,
after the Compiler Agent reports a successful compilation, the Translation
Orchestrator independently runs \texttt{cargo check} to confirm the verdict.
This dual-layer approach guards against agent hallucination and ensures that
pipeline state reflects ground truth rather than agent self-report.
\section{Experimental Setup}
\label{sec:experimentalSetup} 

\begin{table}[t]
\centering
\caption{Benchmark statistics. Projects are sampled from the
CRUST-Bench collection of 100 C repositories.}
\label{tab:benchmark}
\footnotesize
\setlength{\tabcolsep}{4pt}
\renewcommand{\arraystretch}{0.98}
\begin{tabular}{lrrrr}
\toprule
\textbf{Project} & \textbf{LoC} & \textbf{\#Func} & \textbf{\#Files} & \textbf{\#Tests} \\
\midrule
CB        &   213 &  11 &  2 &   18 \\
S-Conf    &   719 &  51 &  4 &   41 \\
VaultSync & 1,121 &  41 & 20 &    4 \\
mvptree   & 1,121 &  27 &  2 &   26 \\
RazzSim   & 1,145 &  45 &  5 &  138 \\
Remimu    & 1,162 &   3 &  1 &    3 \\
libpgn    & 1,162 &  64 & 26 &  476 \\
LTRE      & 1,212 &  45 &  4 &  368 \\
LC-Eval   & 1,264 &  70 & 15 &  101 \\
jccc      & 1,310 &  60 & 22 &   23 \\
libpsbt   & 1,331 &  60 & 11 &    9 \\
mdb       & 1,340 &  28 &  4 &   52 \\
GNNSC     & 1,367 &  65 & 28 &   24 \\
impcheck  & 1,397 & 105 & 27 &   83 \\
cfsm      & 1,446 &   4 & 13 &    5 \\
worsp     & 1,494 &  62 &  2 &   70 \\
libutf    & 1,525 &  81 & 33 &   87 \\
KComp     & 1,589 & 136 & 12 &    2 \\
fslib     & 1,958 & 103 & 24 &   11 \\
XOpt      & 2,113 &  25 &  3 &   18 \\
RecMgr    & 2,400 &  82 & 16 &   24 \\
libm17    & 2,901 &  45 & 10 &   26 \\
tisp      & 3,562 & 100 &  9 &  378 \\
Megalania & 3,621 & 101 & 41 &   27 \\
\bottomrule
\end{tabular}
\noindent\begin{minipage}{\columnwidth}
\vspace{0.3em}
\emph{Name Abbreviations:}
CB: CircularBuffer;
S-Conf: Simple-Config;
RazzSim: razz\_simulation;
LC-Eval: lambda-calculus-eval;
GNNSC: Genetic-neural-network-for-simple-control;
KComp: kairoCompiler;
RecMgr: recordManager.
\end{minipage}
\end{table}

\textbf{Benchmark.} 
Table~\ref{tab:benchmark} summarizes key characteristics of the selected benchmarks, including
lines of code (LoC), number of functions (\#Func), number of files (\#Files),
and the number of test cases. Our primary dataset is based on
CRUST-Bench~\cite{khatry2025crust}, a benchmark of 100 C repositories paired with
manually generated Rust interfaces. We extract benchmark statistics directly from the original dataset, applying minor corrections to discrepancies in the reported test case counts. We evaluate \orbit{} on a subset of 24
programs, prioritizing repositories larger than 1,000 LoC and
including a few smaller programs to preserve diversity across application types.
We focus on larger repositories because prior work~\cite{farrukh2025safetrans} has shown
that LLM-based translation techniques struggle as program size and structural
complexity increase. This design allows us to evaluate \orbit{} on more
challenging and realistic migration tasks.

CRUST-Bench includes a broad range of real-world applications from several
domains, including programming language infrastructure, algorithmic libraries,
and system utilities. 

\noindent\textbf{Agents and LLMs.}
To demonstrate the extensibility and adaptability of \orbit{}, we employ two
off-the-shelf coding agents: Opencode~\cite{OpenCodeAI} and Codex~\cite{OpenAICodex}.
Opencode is an open-source coding agent compatible with over 75 LLM providers,
making it a highly versatile choice. We configure it with Qwen3-Coder-480B, an
open-source model specialized for coding tasks, accessed via AWS Bedrock. While
Codex (from OpenAI) supports both local LLMs and proprietary models, our
preliminary testing revealed frequent silent failures and tool-calling issues
when utilizing non-OpenAI models. We therefore use Codex with the GPT-5.2-Codex
model.


\section{Experimental Evaluation}
\label{sec:results}

\newcommand{\na}{--}
\begin{table*}[t]
\centering
\caption{Correctness results on the 24 CRUST-Bench programs.
For ORBIT, we distinguish two interface modes:
\orbitm{} uses expert-written Rust interfaces
provided by CRUST-Bench, while \orbita{} uses
interfaces generated automatically by \orbit{}.
\cmark~= success; \xmark~= failure;
\na~= not applicable.
Abbreviated names: \textbf{CB}~= CircularBuffer;
\textbf{S-Conf}~= Simple-Config;
\textbf{RazzSim}~= razz\_simulation;
\textbf{LC-Eval}~= lambda-calculus-eval;
\textbf{GNNSC}~= Genetic-neural-network-for-simple-control;
\textbf{KComp}~= kairoCompiler;
\textbf{RecMgr}~= recordManager.
}
\label{tab:rq1Correctness}
\footnotesize
\setlength{\tabcolsep}{7pt}
\renewcommand{\arraystretch}{0.98}
\begin{tabular}{lrccccccccc}
\toprule
& & \multicolumn{3}{c}{\textbf{C2Rust}}
  & \multicolumn{2}{c}{\textbf{CRUST-Bench}}
  & \multicolumn{2}{c}{\textbf{ORBIT\textsubscript{Ext}}}
  & \multicolumn{2}{c}{\textbf{ORBIT\textsubscript{Gen}}} \\
\cmidrule(lr){3-5}\cmidrule(lr){6-7}\cmidrule(lr){8-9}\cmidrule(l){10-11}
\textbf{Project} & \textbf{LoC}
  & \makecell{\textbf{Trans.}\\\textbf{Success}}
  & \makecell{\textbf{Comp.}\\\textbf{Success}}
  & \makecell{\textbf{Test Success}}
  & \makecell{\textbf{Comp.}\\\textbf{Success}}
  & \makecell{\textbf{Test Success}}
  & \makecell{\textbf{Comp.}\\\textbf{Success}}
  & \makecell{\textbf{Test Success}}
  & \makecell{\textbf{Comp.}\\\textbf{Success}}
  & \makecell{\textbf{Test Success}} \\
\midrule
CB        &   213 & \cmark & \cmark & \cmark & \cmark & \cmark & \cmark & \cmark & \cmark & \cmark \\
S-Conf    &   719 & \xmark & \na & \na    & \cmark & \xmark & \cmark & \cmark & \cmark & \cmark \\
VaultSync & 1,121 & \cmark & \xmark & \na    & \xmark & \na    & \cmark & \cmark & \cmark & \cmark \\
mvptree   & 1,121 & \cmark & \xmark & \na    & \cmark & \cmark & \cmark & \cmark & \cmark & \cmark \\
RazzSim   & 1,145 & \cmark & \cmark & \xmark & \cmark & \xmark & \cmark & \cmark & \cmark & \cmark \\
Remimu    & 1,162 & \cmark & \cmark & \cmark & \xmark & \na    & \cmark & \cmark & \cmark & \cmark \\
libpgn    & 1,162 & \cmark & \cmark & \xmark & \cmark & \xmark & \cmark & \xmark & \cmark & \cmark \\
LTRE      & 1,212 & \cmark & \xmark & \na    & \cmark & \cmark & \cmark & \cmark & \cmark & \xmark \\
LC-Eval   & 1,264 & \cmark & \cmark & \cmark & \cmark & \cmark & \cmark & \cmark & \cmark & \cmark \\
jccc      & 1,310 & \cmark & \cmark & \xmark & \xmark & \na    & \cmark & \cmark & \cmark & \cmark \\
libpsbt   & 1,331 & \cmark & \cmark & \xmark & \xmark & \na    & \cmark & \cmark & \cmark & \cmark \\
mdb       & 1,340 & \cmark & \xmark & \na    & \cmark & \cmark & \cmark & \cmark & \cmark & \cmark \\
GNNSC     & 1,367 & \cmark & \xmark & \na    & \xmark & \na    & \cmark & \cmark & \cmark & \cmark \\
impcheck  & 1,397 & \xmark & \na    & \na    & \xmark & \na    & \cmark & \xmark & \cmark & \cmark \\
cfsm      & 1,446 & \cmark & \cmark & \cmark & \cmark & \xmark & \cmark & \cmark & \cmark & \cmark \\
worsp     & 1,494 & \cmark & \cmark & \xmark & \xmark & \na    & \cmark & \cmark & \cmark & \cmark \\
libutf    & 1,525 & \cmark & \cmark & \xmark & \xmark & \na    & \cmark & \cmark & \cmark & \cmark \\
KComp     & 1,589 & \cmark & \cmark & \xmark & \xmark & \na    & \cmark & \cmark & \cmark & \cmark \\
fslib     & 1,958 & \cmark & \cmark & \xmark & \xmark & \na    & \cmark & \cmark & \cmark & \cmark \\
XOpt      & 2,113 & \cmark & \xmark & \na    & \cmark & \cmark & \cmark & \cmark & \cmark & \cmark \\
RecMgr    & 2,400 & \xmark & \na    & \na    & \xmark & \na    & \cmark & \cmark & \cmark & \cmark \\
libm17    & 2,901 & \cmark & \cmark & \xmark & \cmark & \xmark & \cmark & \cmark & \cmark & \cmark \\
tisp      & 3,562 & \cmark & \cmark & \xmark & \xmark & \na    & \cmark & \cmark & \cmark & \xmark \\
Megalania & 3,621 & \cmark & \cmark & \cmark & \xmark & \na    & \cmark & \cmark & \cmark & \cmark \\
\midrule
\textbf{Success Rate} &
  & \makecell{21/24\\(87.5\%)} & \makecell{15/24\\(62.5\%)} & \makecell{5/24\\(20.8\%)}
  & \makecell{11/24\\(45.8\%)} & \makecell{6/24\\(25.0\%)}
  & \makecell{\textbf{24/24}\\\textbf{(100\%)}} & \makecell{\textbf{22/24}\\\textbf{(91.7\%)}}
  & \makecell{\textbf{24/24}\\\textbf{(100\%)}} & \makecell{\textbf{22/24}\\\textbf{(91.7\%)}} \\
\bottomrule
\end{tabular}
\end{table*}

In this section we present the results of our experimental evaluation,
focusing on the following main research questions:
\begin{itemize}
    \item RQ1: Effectiveness of \orbit{} relative to existing approaches in terms of correctness and safety.
    \item RQ2: Evaluation using the DARPA TRACTOR dataset.
    \item RQ3: Evaluation of individual \orbit{} components towards translation success.
\end{itemize}

\subsection{RQ1: Correctness and Safety}

\subsubsection{Correctness Evaluation}

Table~\ref{tab:rq1Correctness} presents the correctness results of 
\orbit{} alongside two baselines on compilation and test success metrics:
i)~\emph{C2Rust}~\cite{immunant2022c2rust}: a rule-based 
    transpiler developed by Immunant and Galois that 
    converts C code to Rust without LLM assistance;
ii)~\emph{CRUST-Bench}~\cite{khatry2025crust}: a multi-step 
    self-repair framework for project-level C-to-Rust translation that 
    employs a repair feedback loop to automatically correct 
    translation errors.
The bottom row of the table reports the overall success rate across 
all evaluated tools.


\noindent\textbf{C2Rust.}
C2Rust successfully transpiles 21 out of 24 programs (87.5\%), but its 
compilation success rate drops sharply to 15/24 (62.5\%), revealing 
that successful transpilation alone does not guarantee compilable output. Among the three 
transpilation failures, one originates from a fundamental limitation of its Clang-based
frontend: C2Rust cannot process GCC nested functions, a non-standard GCC extension~\cite{gcc_nested_functions} that permits function definitions to appear inside other function bodies.
In \texttt{recordManager}, the function \texttt{openTable} defines two
nested helpers (\texttt{readIntFromHeader} and
\texttt{setSchemaAttributes}) directly within its body
(Listing~\ref{lst:nested_c}).
Since Clang rejects this GCC extension, C2Rust aborts at the
transpilation stage rather than producing incorrect output.

\orbit{} resolves this by restructuring the nested functions as a Rust 
closure captured within the enclosing function's scope 
(Listing~\ref{lst:nested_rs}). The closure \texttt{read\_i32} captures 
a mutable reference to the byte offset, producing semantically 
equivalent behaviour while remaining fully safe Rust with no 
\texttt{unsafe} blocks. This example illustrates a broader advantage 
of agentic translation over rule-based transpilation: the LLM agent is 
not constrained by the syntactic subset of C accepted by a particular 
compiler frontend, enabling it to
produce idiomatic restructurings.

\begin{figure}[t]
\begin{lstlisting}[language=C, frame=single,
  caption={GCC nested functions inside \texttt{openTable} 
  (\texttt{recordManager.c}).},
  label={lst:nested_c}]
// Nested functions defined inside openTable()
int readIntFromHeader(char **hdr) {
  int val = *(int *)(*hdr);
  *hdr += sizeof(int);
  return val;
}
void setSchemaAttributes(
    Schema *s, char **hdr) {
  s->numAttr = readIntFromHeader(hdr);
  s->keySize = readIntFromHeader(hdr);
}
\end{lstlisting}

\vspace{4pt}

\begin{lstlisting}[language=Rust, frame=single,
  caption={\orbit{} restructures nested functions as a closure 
  (Rust).},
  label={lst:nested_rs}]
// Closure inside open_table() captures mutable offset
let mut read_i32 =
  |data: &[u8], off: &mut usize| -> i32 {
    let v = i32::from_le_bytes(
      data[*off..*off + 4]
        .try_into().unwrap()
    );
    *off += 4;
    v
  };
schema.num_attr = read_i32(header, &mut offset);
schema.key_size = read_i32(header, &mut offset);
\end{lstlisting}
\end{figure}

Even among the 15 programs that compile successfully, C2Rust achieves 
a test success rate of only 5/24 (20.8\%), with runtime crashes being the
major reason.


\noindent\textbf{CRUST-Bench.}
We run the CRUST-Bench pipeline with Qwen3-Coder-480B under the 
default setting of three self-repair rounds per task and a 
single-candidate greedy repair strategy, as the authors report that 
these configurations yield the best results~\cite{khatry2025crust}. 
CRUST-Bench achieves a compilation success rate of 11/24 (45.8\%) and 
a test success rate of 6/24 (25.0\%). The translations frequently 
contain \texttt{unimplemented!} macro invocations or incomplete fragments---a known limitation 
also reported by the original authors---that prevent successful end-to-end execution. By contrast, 
\orbit{}'s fine-grained control over the translation process eliminates 
such failures entirely, achieving 100\% compilation success on all 24 
programs.


\begin{table*}[t]
\centering
\caption{Safety analysis comparison across translation tools.
\textbf{ptr\_d} = raw pointer declarations;
\textbf{ptr\_*} = raw pointer dereferences;
\textbf{uLoC} = unsafe lines of code;
\textbf{\%Unsafe} = unsafe percentage of total LoC;
\textbf{LoC} = total lines of code.
``\na{}'' indicates the project was either not translated or did not compile successfully for that tool.
Highlighted rows (\colorbox{hlrow}{\phantom{xx}}) denote programs
where at least one ORBIT configuration produced non-zero unsafe code.
}
\label{tab:safety}
\scriptsize
\setlength{\tabcolsep}{2pt}
\resizebox{\linewidth}{!}{%
\renewcommand{\arraystretch}{0.98}
\begin{tabular}{lrrrrrrrrrrrrrrrrrrrrr}
\toprule
& \multicolumn{5}{c}{\textbf{C2Rust}}
& \multicolumn{5}{c}{\textbf{CRUST-Bench}}
& \multicolumn{5}{c}{\textbf{ORBIT\textsubscript{Ext}}}
& \multicolumn{5}{c}{\textbf{ORBIT\textsubscript{Gen}}} \\
\cmidrule(lr){2-6}\cmidrule(lr){7-11}\cmidrule(lr){12-16}\cmidrule(l){17-21}
\textbf{Project}
  & \textbf{ptr\_d} & \textbf{ptr\_*} & \textbf{uLoC} & \textbf{\%Unsafe} & \textbf{LoC}
  & \textbf{ptr\_d} & \textbf{ptr\_*} & \textbf{uLoC} & \textbf{\%Unsafe} & \textbf{LoC}
  & \textbf{ptr\_d} & \textbf{ptr\_*} & \textbf{uLoC} & \textbf{\%Unsafe} & \textbf{LoC}
  & \textbf{ptr\_d} & \textbf{ptr\_*} & \textbf{uLoC} & \textbf{\%Unsafe} & \textbf{LoC} \\
\midrule
CB        & 171  & 100  & 530  & 82.3 & 644   & 0   & 0 & 0  & 0.0 & 292  & 0  & 0 & 0  & 0.0 & 273  & 0  & 0 & 0  & 0.0 & 259  \\
S-Conf    & \na  & \na  & \na & \na & \na  & 0   & 0 & 0  & 0.0 & 791  & 0  & 0 & 0  & 0.0 & 1074 & 0  & 0 & 0  & 0.0 & 1646 \\
VaultSync & \na & \na & \na & \na & \na  & \na & \na & \na & \na & \na & 0 & 0 & 0 & 0.0 & 959  & 0  & 0 & 0  & 0.0 & 1484 \\
mvptree   & \na & \na & \na & \na & \na   & 0   & 0 & 0  & 0.0 & 653  & \hl{0} & \hl{0} & \hl{1}  & \hl{0.1} & \hl{1147} & 0  & 0 & 0  & 0.0 & 2153 \\
RazzSim   & 391  & 211  & 1204 & 67.4 & 1787  & 2   & 0 & 6  & 1.0 & 607  & 0  & 0 & 0  & 0.0 & 1285 & \hl{0}  & \hl{0} & \hl{9}  & \hl{0.6} & \hl{1572} \\
Remimu    & 1108 & 718  & 5258 & 97.4 & 5399  & \na & \na & \na & \na & \na & 0 & 0 & 0 & 0.0 & 1318 & 0  & 0 & 0  & 0.0 & 1087 \\
libpgn    & 513  & 421  & 2006 & 75.0 & 2673  & 0   & 0 & 0  & 0.0 & 1246 & 0  & 0 & 0  & 0.0 & 2902 & 0  & 0 & 0  & 0.0 & 1952 \\
LTRE      & \na & \na & \na & \na & \na & 0   & 0 & 0  & 0.0 & 1353 & 0  & 0 & 0  & 0.0 & 2827 & 0  & 0 & 0  & 0.0 & 1107 \\
LC-Eval   & 1004 & 350  & 2235 & 68.5 & 3264  & 0   & 0 & 26 & 2.5 & 1023 & 0  & 0 & 0  & 0.0 & 1501 & 0  & 0 & 0  & 0.0 & 1671 \\
jccc      & 1267 & 266  & 2131 & 64.2 & 3321  & \na & \na & \na & \na & \na & 0 & 0 & 0 & 0.0 & 1662 & 0  & 0 & 0  & 0.0 & 1849 \\
libpsbt   & 645  & 328  & 1920 & 75.3 & 2550  & \na & \na & \na & \na & \na & 0 & 0 & 0 & 0.0 & 1535 & 44 & 0 & 0  & 0.0 & 1578 \\
mdb       & \na & \na & \na & \na & \na & 0   & 0 & 0  & 0.0 & 560  & 0  & 0 & 0  & 0.0 & 784  & 0  & 0 & 0  & 0.0 & 1596 \\
GNNSC     & \na & \na & \na & \na & \na   & \na & \na & \na & \na & \na & 0 & 0 & 0 & 0.0 & 2145 & 0  & 0 & 0  & 0.0 & 2116 \\
impcheck  & \na  & \na  & \na  & \na   & \na   & \na & \na & \na & \na & \na & \hl{1} & \hl{0} & \hl{2} & \hl{0.1} & \hl{2193} & 0  & 0 & 0  & 0.0 & 2839 \\
cfsm      & 361  & 105  & 2203 & 86.0 & 2563  & 0   & 0 & 0  & 0.0 & 75   & 0  & 0 & 0  & 0.0 & 195  & 0  & 0 & 0  & 0.0 & 334  \\
worsp     & 696  & 879  & 2398 & 81.7 & 2936  & \na & \na & \na & \na & \na & 0 & 0 & 0 & 0.0 & 2032 & 0  & 0 & 0  & 0.0 & 3581 \\
libutf    & 368  & 382  & 2441 & 82.8 & 2950  & \na & \na & \na & \na & \na & 0 & 0 & 0 & 0.0 & 1656 & 1  & 0 & 0  & 0.0 & 2073 \\
KComp     & 952  & 236  & 2021 & 62.1 & 3253  & \na & \na & \na & \na & \na & \hl{8} & \hl{1} & \hl{4}  & \hl{0.2} & \hl{1997} & \hl{10} & \hl{0} & \hl{12} & \hl{0.7} & \hl{1668} \\
fslib     & 1269 & 734  & 4264 & 71.9 & 5928  & \na & \na & \na & \na & \na & 0  & 0 & 0  & 0.0 & 3412 & \hl{3}  & \hl{6} & \hl{45} & \hl{1.4} & \hl{3197} \\
XOpt      & \na  & \na  & \na & \na & \na  & 6   & 0 & 0  & 0.0 & 837  & \hl{20} & \hl{4} & \hl{16} & \hl{1.0} & \hl{1591} & 0  & 0 & 0  & 0.0 & 1789 \\
RecMgr    & \na  & \na  & \na  & \na   & \na   & \na & \na & \na & \na & \na & \hl{2} & \hl{0} & \hl{4}  & \hl{0.2} & \hl{2450} & 0  & 0 & 0  & 0.0 & 3075 \\
libm17    & 240  & 242  & 1446 & 58.5 & 2473  & 0   & 0 & 49 & 4.0 & 1239 & 0  & 0 & 0  & 0.0 & 1141 & 0  & 0 & 0  & 0.0 & 1309 \\
tisp      & 1626 & 1189 & 5644 & 20.4 & 27741 & \na & \na & \na & \na & \na & 0 & 0 & 0 & 0.0 & 3341 & 0  & 0 & 0  & 0.0 & 6906 \\
Megalania & 1362 & 612  & 4186 & 51.2 & 8181  & \na & \na & \na & \na & \na & 4 & 0 & 0 & 0.0 & 1725 & 0  & 0 & 0  & 0.0 & 2690 \\
\bottomrule
\end{tabular}}%
\end{table*}

\noindent\textbf{ORBIT.}
As described in Section~\ref{sec:methodology}, \orbit{} supports both 
manually provided and automatically generated Rust interfaces. To 
evaluate each capability independently, we run \orbit{} in two 
configurations: \orbitm{} and \orbita{}.
The former uses expert-written Rust interfaces provided by CRUST-Bench.
The latter employs 
Codex with GPT-5.2-Codex during the Agentic Iterative Scaffolding 
phase for interface generation, and it then switches to Opencode with 
Qwen3-Coder-480B during translation orchestration, balancing 
translation quality against 
cost.\footnote{gpt-5.2-codex is significantly more expensive 
than qwen: input costs are approximately $3.89\times$ 
higher (\$1.75 vs.\ \$0.45 per 1M tokens) and output costs are 
approximately $7.78\times$ higher (\$14.00 vs.\ \$1.80 per 1M 
tokens)~\cite{OpenAIPricing2026, AWSBedrockPricing2026}.}

\orbitm{} achieves 100\% compilation success and a test success rate of 22/24 
(91.7\%). \orbita{}, which generates interfaces automatically, matches the 
compilation success of \orbitm{} at 24/24 (100\%), and achieves an 
identical test success rate of 22/24 (91.7\%), with failures on 
\texttt{LTRE} and \texttt{tisp} instead. The near-identical performance 
between the two modes demonstrates that \orbit{}'s automatic interface 
generation is a viable substitute for manually crafted interfaces, 
removing a significant practical barrier to adoption.

Compared to both baselines, \orbit{} substantially improves end-to-end
correctness. Relative to C2Rust, \orbit{} improves compilation success from
58.3\% to 100\% and test-suite success from 20.8\% to 91.7\%. Relative to
CRUST-Bench, it improves compilation success from 45.8\% to 100\% and
test-suite success from 25.0\% to 91.7\%. These improvements suggest that the
main gains of \orbit{} come from its modular and dependency aware agentic approach.


\subsubsection{Memory Safety Evaluation}

Table~\ref{tab:safety} reports the memory safety results across all
four tools. We measure safety as the percentage of unsafe lines of code
(\%Unsafe) in the translated Rust output, with lower values indicating
a safer translation. Unlike prior studies~\cite{khatry2025crust, li2025adversarial, farrukh2025safetrans} that often perform aggregated memory safety analysis, \orbit{} adopts a more granular, multi-dimensional safety taxonomy:
number of raw pointer declarations and dereferences, along with total number
of unsafe lines of code in relation to the total LoC, which complements the
percentage unsafety.
A dash  (``\na{}'') in the table indicates either that the tool fails the transpilation process, or its translation contains build failures.

\noindent\textbf{Baselines.}
C2Rust produces translations that are unsafe throughout, averaging 69.6\% of unsafe lines of code across the 15 programs it successfully compiles, ranging
from 20.4\% (\texttt{tisp}) to 97.4\% (\texttt{Remimu}). This is
expected, as C2Rust performs a structural one-to-one mapping of C
constructs to Rust, preserving raw pointer arithmetic verbatim rather
than reasoning about ownership. Since CRUST-Bench translations include manually generated Rust Interfaces, this assists the iterative repair loop to
reduce unsafe code blocks substantially, averaging 0.68\% across the 11
programs it translates. However, three programs retain non-zero lines of unsafe
code even after using safe interfaces: \texttt{razz\_simulation} (1\%),
\texttt{lambda-calculus-eval} (2.5\%), and \texttt{libm17}
(4\%). Despite the claim that CRUST-Bench provides safe interfaces, we
observed some instances of \texttt{static mut} usage in the source code that
limit the LLM's ability to generate safe code. For example, \texttt{libm17} contains
these global variables in both the original and the translated Rust version. For other
programs with unsafe usage, the root cause varies between
global variables (introduced by
the LLM later), raw pointer casting, or unsafe function calls.
On the 6 programs successfully compiled by all four tools, C2Rust averages 73.0\% of unsafe lines of code, compared to 1.25\% for CRUST-Bench, 0.00\% for \orbitm{}, and 0.10\% for \orbita{}---a reduction of 100\% relative to C2Rust for \orbitm{}, and over 98\% relative to CRUST-Bench.

\noindent\textbf{\orbitm{}.}
\orbitm{} achieves zero lines of unsafe code in 19/24 programs, with an
average \%Unsafe of just 0.06\%. Among the five programs with non-zero
unsafe lines of code, the absolute counts are negligible: \texttt{mvptree}, and
\texttt{impcheck}  contain at most 2 unsafe lines
each, while \texttt{KComp}, \texttt{RecMgr}, and \texttt{XOpt} account for the majority
with maximum 16 unsafe lines, out of translation outputs
exceeding 1,500~LoC. The raw pointer counts reflect the same picture:
\orbitm{} introduces only 35 raw pointer declarations and 5
dereferences across all 24 programs, confirming that the agent
consistently eliminates the pervasive pointer arithmetic of the
original C program and confines any residual unsafe code blocks to isolated locations.
 
\noindent\textbf{\orbita{}.}
\orbita{} achieves a comparable safety profile, achieving zero lines of unsafe
code in 21/24 programs, with an average \%Unsafe of 0.11\%.
\orbita{} eliminates unsafe code in two additional
programs compared to \orbitm{}, though the programs that retain
residual unsafe lines are different (\texttt{RazzSim} (9 lines), \texttt{KComp}
(12 lines), and \texttt{fslib} (45 lines)), suggesting that automatic
interface generation influences which pointer patterns the agent
restructures rather than increasing overall unsafety. Raw pointer
totals remain similarly low at 58 declarations and 6 dereferences.
The near-identical profile between the two configurations confirms
that automatic interface generation does not compromise safety, making
\orbita{} a practical choice when manually written interfaces are
unavailable.

\subsection{RQ2: DARPA TRACTOR Dataset}

\begin{table}[t]
  \centering
  \footnotesize
  \setlength{\tabcolsep}{6pt}
  \caption{Performance of \orbit{} on the hardest TRACTOR programs. 
  Perf. Pass Rate denotes the number of performers (out of six) that successfully translated each case. 
  ORBIT Result shows case-level outcome, and ORBIT Vec. Pass Rate denotes the percentage of test vectors passed.}
  \label{tab:darpaOrbit}
  \renewcommand{\arraystretch}{0.95}
  \begin{tabular}{l l c c c}
  \toprule
  \textbf{Type} & \textbf{Test Case} &
  \makecell{\textbf{Perf.}\\\textbf{Pass Rate}} &
  \makecell{\textbf{ORBIT}\\\textbf{Result}} &
  \makecell{\textbf{ORBIT Vec.}\\\textbf{Pass Rate}} \\
  \midrule
  
  \multirow{5}{*}{Exec}
   & 016\_switch-arith     & 3/6 & Pass    & 100\% \\
   & 042\_float\_union     & 3/6 & Pass    & 100\% \\
   & 033\_bitfield         & 3/6 & Pass    & 100\% \\
   & 030\_int\_underflow   & 2/6 & Pass   & 100\% \\
   & 002\_stdin\_echo      & 3/6 & Partial & 75\% \\
  \midrule
  
  \multirow{8}{*}{Lib}
   & read\_scalefactors\_lib & 3/6 & Pass    & 100\% \\
   & 004\_loop\_lib          & 2/6 & Pass    & 100\% \\
   & read\_side\_info\_lib   & 4/6 & Pass    & 100\% \\
   & wcscat\_lib             & 4/6 & Pass    & 100\% \\
   & update\_frame\_header\_lib & 4/6 & Pass    & 100\% \\
   & 030\_int\_underflow\_lib & 2/6 & Fail    & 0\% \\
   & contrast\_ratio\_lib     & 3/6 & Partial & 62.5\% \\
   & hex2bin\_lib             & 2/6 & Fail    & 0\% \\
  \bottomrule
  \end{tabular}
  \end{table}

DARPA launched the TRACTOR program as a major research effort to
develop scalable, automated techniques for translating large C
codebases into memory-safe Rust~\cite{DARPA_TRACTOR}. As part of the
program, a test battery of 150 C programs was publicly released to evaluate
the translation tools developed by the participating performers
across a range of C features, accompanied by a detailed
evaluation report~\cite{TRACTOR_Report_2024} covering correctness and safety.
We evaluate \orbit{} on this standard benchmark to understand how it
performs compared to the participants' tools, and where it stands on
challenging real-world programs.

The TRACTOR report presents aggregated results across performers,
covering overall failure rates and failure types, rather than
fine-grained per-program, per-performer breakdowns. While this is
sufficient to identify overall trends, it does not reveal which
performer failed a given program or how many test vectors were missed,
preventing a complete per-system comparison at the level of individual
programs and vectors. We therefore focus on the subset of the most
failure-prone programs identified in the evaluation. Since the TRACTOR
test harness requires translations to be in a specific output format,
we extended \orbit{} to support this format. From the top-15 hardest
cases, we select 13 programs, excluding two long-running cases that
require additional configuration changes. The selected set covers both
executable and library programs, allowing us to evaluate \orbit{}
under different interface and behavioural constraints. Since TRACTOR
does not release test vectors to translation systems, and \orbit{}
requires seed tests for its iterative repair process, we construct
initial test vectors manually to enable translation and ensure a fair
comparison. Importantly, all final results are computed using the
official TRACTOR-provided test vectors, which remain completely hidden
from \orbit{} during the translation process.

Table~\ref{tab:darpaOrbit} summarizes the performance of \orbit{}
across all selected programs. Overall, \orbit{} achieves full
correctness (``Pass'') on 9 out of the 13 programs (69.2\%), and at least
partial correctness on 11 of them (84.6\%). When broken
down by program type, \orbit{} attains an 80\% success rate on
executable programs (4/5) and 62.5\% on library programs (5/8).

Despite the lack of per-performer results in the TRACTOR report, a
comparison at the aggregate level reveals that \orbit{} performs
competitively with the six performers. The report identifies
the average performer pass rate across Battery~01 as
82.2\%~\cite{TRACTOR_Report_2024}, with the top performer
reaching 98.7\% and the lowest
reaching 48.0\%. On the 13 programs we evaluated, the TRACTOR baseline pass
rate (i.e., the fraction of performers that successfully translated each
case) ranges from 2/6 to 4/6, confirming these are among the most
challenging programs in the battery. Against this backdrop, \orbit{}
achieves a pass rate of $\sim$70\% (9/13), which falls within the
range of performers on the full battery and notably surpasses the
lowest-performing system on our selected subset. On the executable
programs specifically, \orbit{} matches or exceeds the majority of
performers: all five executable cases have a performer pass rate of
only 2--3 out of 6, yet \orbit{} passes four of them outright and
achieves 75\% vector coverage on the fifth. Among the library
programs, \orbit{} fully passes five cases, including
\texttt{read\_scalefactors\_lib} and \texttt{004\_loop\_lib}, 
of which only 2 out of 6 performers translated \texttt{004\_loop\_lib} successfully, and 3 out of 6 translated \texttt{read\_scalefactors\_lib}.
To enable testing, the
TRACTOR test harness imposes stricter requirements on ABI
compatibility, state preservation, and FFI boundaries. However,
\orbit{} is primarily designed and optimized to avoid the use of FFI
and to perform translation testing using \texttt{cargo test}. This
explains the comparatively lower success rate on library programs
relative to executable programs.

Among the partial and failing cases, the root cause is often not a
failure to model pointer semantics or Rust's ownership system, but
rather subtle behavioral differences between C standard library
functions and their closest Rust equivalents. In
\texttt{contrast\_ratio\_lib}, \orbit{} selected
\texttt{f32::powf} as the counterpart to C's \texttt{pow()}, but C
implicitly promotes \texttt{float} arguments to \texttt{double} before
the computation, performing the exponentiation at 64-bit precision
before casting back. Rust's \texttt{f32::powf} operates entirely in
single precision, producing slightly different results on
boundary-value inputs. Similarly, in \texttt{002\_stdin\_echo}, \texttt{write\_all} is used as the natural counterpart to
\texttt{fputs}, yet \texttt{fputs} silently truncates the output at the
first null byte whereas \texttt{write\_all} writes the full byte
slice. These failures highlight a remaining gap in LLM knowledge of
the precise behavioral contracts of C functions and the subtle ways
their Rust counterparts diverge from them---a limitation of current
LLM-based approaches that calls for deeper exploration.

\subsection{RQ3: Ablation Study}
\label{sec:ablation}

To evaluate the contribution of the key components of \orbit{} in overall
translation success, we conduct an
ablation study across four settings:
i)~\texttt{full}, which runs the complete
pipeline;
ii)~\texttt{base}, which translates the entire project in a single prompt
without the dependency graph, Rust interfaces, and translation orchestrator;
iii)~\texttt{w/o interfaces}, which retains the translation orchestrator and
dependency graph but removes the initial Rust scaffold and function mappings,
requiring the agent to construct the Rust project structure from scratch; and
iv)~\texttt{w/o mapping}, which retains both the dependency graph and Rust
interfaces but removes explicit function-level target mappings, leaving the
agent to identify Rust counterparts independently.
All settings use the same
Opencode agent with the Qwen3-Coder-480B model. We perform our evaluation on three benchmarks
of varying size and complexity: \texttt{CircularBuffer} (213~LoC, 11 functions),
\texttt{LTRE} (1,212~LoC, 45 functions), and \texttt{libm17} (2,901~LoC, 45
functions). Beyond build and test success, which can give a false sense of
correctness, we introduce two additional metrics:
 \textbf{(i) Functional Coverage (\%)}, defined as the percentage of C functions fully translated into Rust (excluding stubs, \texttt{unimplemented!()} placeholders, and missing functions), and \textbf{(ii) Test Coverage (\%)}, defined as the percentage of original C test cases translated into the Rust test suite and executable via \texttt{cargo test}.
These metrics are intentionally stricter than build and test success alone, as
our manual audit revealed multiple cases where a project was built and tested
successfully despite missing functionality, shallow test coverage, or translated
test harnesses that were never executed.
Table~\ref{tab:ablation} reports our results across all above settings.

\begin{table}[t]
  \centering
  \footnotesize
  \setlength{\tabcolsep}{4.0pt}
  \caption{Ablation results for \textsc{Orbit} on three benchmarks.
    Func.\ Coverage is the percentage of C functions fully implemented
    in Rust. Test Coverage is measured against the original C test
    logic. Build success denotes passing \texttt{cargo check}; Test
    success denotes passing \texttt{cargo test}.
    (CB~= CircularBuffer)}
  \label{tab:ablation}
\renewcommand{\arraystretch}{0.98}
  \begin{tabular}{ll cccc}
  \toprule
  \textbf{Prog.} & \textbf{Mode}
    & \makecell{\textbf{Func.}\\\textbf{Coverage} \textbf{(\%)}}
    & \makecell{\textbf{Test}\\\textbf{Coverage} \textbf{(\%)}}
    & \makecell{\textbf{Build}\\\textbf{Succ.}}
    & \makecell{\textbf{Test}\\\textbf{Succ.}} \\
  \midrule
  \multirow{4}{*}{CB}
  & base            & 100.0 &   0.0               & \cmark & \cmark \\
  & w/o interfaces  & 100.0 &  90.9               & \cmark & \cmark \\
  & w/o mapping     & 100.0 & $\na^{*}$           & \cmark & \cmark \\
  & full            & 100.0 & $\na^{*}$           & \cmark & \cmark \\
  \midrule
  \multirow{4}{*}{LTRE}
  & base            &  64.4 &   3.0               & \cmark & \cmark \\
  & w/o interfaces  &  82.2 &  12.2               & \cmark & \cmark \\
  & w/o mapping     &  93.3 & $\na$               & \cmark & \cmark \\
  & full            & 100.0 & $\na$               & \cmark & \xmark \\
  \midrule
  \multirow{4}{*}{libm17}
  & base            &  84.1 &   0.0               & \xmark & \xmark \\
  & w/o interfaces  & 100.0 & 100.0               & \cmark & \cmark \\
  & w/o mapping     & 100.0 & $\na$               & \cmark & \cmark \\
  & full            & 100.0 & $\na$               & \cmark & \cmark \\
  \bottomrule
  \end{tabular}
  \vspace{4pt}
  \par\noindent\scriptsize
  $^{*}$Test Coverage is not applicable for \texttt{w/o mapping} and
  \texttt{full} as the Rust scaffold provides complete test
  implementations.
\end{table}
 
\subsubsection{Impact of Removing All Components (\texttt{Base})} The single-shot
\texttt{base} setting reveals the limitations of vanilla coding agents in
translating an entire project without structural guidance. On
\texttt{CircularBuffer}, where the entire codebase fits within a single context
window, \texttt{base} achieves full functional coverage, compiles successfully
and passes all tests, demonstrating that a single-prompt approach is viable for
small, self-contained programs. However, it produces no translated tests (0.0\%
test coverage), meaning the test success rests entirely on three shallow unit
tests written by the agent rather than a translation of the original C
validation suite. As program complexity increases, \texttt{base} degrades
substantially. On \texttt{LTRE}, functional coverage drops to 64.4\%---seven
functions are missing outright and six are stubs---while test coverage collapses
to 3.0\% (12 Rust tests vs. $\sim$450 C cases), with several of those tests
passing only because they exercise incomplete code paths. On \texttt{libm17},
\texttt{base} fails to compile entirely due to syntax errors, including type mismatches, and borrow checker violations, producing no
tests.
 
\subsubsection{Impact of Removing Interfaces (\texttt{W/O Interfaces})} Removing the
initial Rust scaffold, which in the \texttt{full} \orbit{} setting is constructed
using Agentic Iterative Scaffolding, while retaining the translation orchestrator
that performs incremental translation using the dependency graph, produces mixed
results. For \texttt{CircularBuffer} and \texttt{libm17}, the agent successfully
constructs the project structure from scratch, achieving 100\% functional
coverage and passing all tests. For \texttt{libm17} specifically, this setting
produces the most comprehensive test suite of any non-full mode, as the agent autonomously generates C test vectors to
cross-verify its Rust implementation. This demonstrates that the agent's
self-directed context curation can compensate for the absence of a predefined
scaffold when the project structure is straightforward. However, \texttt{LTRE}
exposes the limits of this compensation: without a predefined interface
structure, the agent leaves four critical functions as stubs or
incomplete, including \texttt{nfa\_clone} and
\texttt{nfa\_uncomplement}, resulting in 82.2\% functional coverage and only
12.2\% test coverage. The absence of a scaffold forces the agent to spend
significant effort on project initialization and macro visibility issues,
leaving less capacity for logic translation.
 
\subsubsection{Impact of Removing Function Mappings (\texttt{W/O Mapping})} Providing
the dependency graph and Rust interfaces but omitting explicit C-to-Rust
function mappings consistently improves functional coverage over \texttt{base}
and \texttt{w/o interfaces} for complex programs. Although \texttt{LTRE} reaches 93.3\%
functional coverage in this setting---the highest of any non-full mode---and
uniquely implements functions absent in both other ablation modes, such as
\texttt{dfa\_serialize} and \texttt{dfa\_deserialize}, \orbit{} still skips some functions; when run in full mode with mapping, \texttt{LTRE} reaches up to 100\% functional coverage. The predefined interface
scaffold provides a stable target structure, allowing the agent to focus on
logic translation rather than architectural decisions. For
\texttt{CircularBuffer} and \texttt{libm17}, the setting achieves full
functional and test coverage.

The contrasting results between \texttt{LTRE} and
\texttt{libm17} across all ablation modes reflect fundamental differences in
their structural complexity rather than program size alone. \texttt{LTRE}, despite being
smaller (1,212 LoC) in size, is structurally more difficult for translation
because its core functionality is organized as a tightly coupled automata
pipeline. Correct behavior depends on interactions among parsing, NFA
construction, DFA compilation, DFA minimization, matching, and round-trip
operations such as serialization and decompilation. As a result, defects in one
stage often invalidate downstream behavior, and local fixes do not necessarily
recover global correctness. In contrast, libm17 is larger (2,901 LoC) but more
decomposable. Many of its components, such as CRC computation, callsign
encoding, and several math helpers, have strong local specifications and can be
validated independently with direct test vectors. This makes iterative
function-level translation more effective despite the larger codebase.

\subsubsection{Full Pipeline} The complete \orbit{} pipeline achieves 100\% functional
and test coverage on all three benchmarks. The one exception is \texttt{LTRE}'s
test success, which fails not due to incomplete translation but due to the
C-to-Rust semantic gap.
To further understand the effectiveness of Function mapping beyond the
mentioned metrics, we conducted a small controlled experiment: we directed a
single coding agent to translate a single function (\texttt{bitset\_get}) from
\texttt{LTRE} in isolation, once with explicit function mapping and once
without. With mapping, the agent completed the translation in 3 tool calls (two
reads and one write). Without mapping, the agent required 6 tool calls, of which
4 were codebase navigation steps to locate the relevant C and Rust files. While
this observation is based on a single function, it illustrates how function
mappings eliminate navigation overhead that would otherwise compound across the
hundreds of per-function translation steps in a full project.
 
Based on the above results, we conclude that
each component of \orbit{} contributes meaningfully to translation quality. The
dependency graph enables incremental translation that scales to complex
projects. Rust interfaces provide the structural scaffold that keeps the agent
focused on logic translation. Removing any single component degrades either
functional or test coverage. The \texttt{base} setting demonstrates that
single-shot translation, while sufficient for small programs, is fundamentally
inadequate for projects beyond a few hundred lines of code.
\section{Threats to Validity} 

Like other LLM-based C-to-Rust translation approaches~\cite{shiraishi2024smartc2rust, sactor}, \orbit{} relies on the tests available with the source code to assess semantic equivalence between the source and translated programs. Consequently, test quality determines the upper bound on the equivalence that can be established between the C and Rust versions. A natural extension would be to harness the capabilities of coding agents to generate additional test cases that expose divergent behaviors, which we leave as future work.
Although \orbit{}'s modular design improves the performance of open-source models and can provide a cost benefit, optimizing the agentic pipeline for cost and token consumption presents a distinct set of challenges orthogonal to the goals of this work and is therefore not our focus. 
Additionally, while \orbit{} operates with full agent autonomy, we observed instances where the agent oscillated between pursuing safe Rust alternatives and completing functionality, continuing until the attempt limit was exhausted. This highlights the need for mechanisms to detect and resolve such agent loops early.
Finally, the non-deterministic nature of LLMs poses a threat to validity, as repeated runs may produce different results. To mitigate this, we adopt a strict correctness criterion: a translation is marked as \textit{Test Success} only if it passes all available test cases.  
\section{Related Work}
\label{sec:related_work}

\textbf{Rule-based C-to-Rust Translation.}
Early automated C-to-Rust translation relied primarily on rule-based
transpilation. C2Rust~\cite{immunant2022c2rust} is the most widely
used tool in this category, applying predefined and custom rewrite
rules to produce Rust code. While it scales to large
codebases, the output is unidiomatic and saturated with
\texttt{unsafe} blocks. Emre et al.~\cite{safer-rust} systematically
characterize the sources of unsafety in C2Rust translations and
propose a compiler-feedback-driven technique to refactor a class of
raw pointers into safe Rust references. Building on C2Rust,
CROWN~\cite{crown} lifts raw pointers to Rust references, though its
scope is restricted to mutable, non-array pointer types. A parallel
line of work targets narrower translation challenges: ConCrat~\cite{hong2023concrat}
focuses on converting lock APIs, while other
tools~\cite{hong2024don, hongc2rust} address specific data type
conversions. Collectively, these approaches demonstrate that
rule-based translation can handle isolated syntactic patterns but
struggles to produce safe, idiomatic Rust at the whole-program level.

\textbf{LLM-based C-to-Rust Translation.}
Recent advances in large language models (LLMs) have enabled their application across a wide range of domains~\cite{HUSEIN2025103917, xu2024large, bozkir2024embedding, kim2025explainablexrunderstandinguser, talha2025lowering, talha2026extending, sapkota2025image, muzammil2025babylon, jelodar2025large}, including code generation and program translation. Pan et al.~\cite{pan2024lost} provide a comprehensive evaluation of LLM-based translation across multiple programming languages, including C-to-Rust, and introduce a taxonomy of common translation errors. To improve translation quality, several works augment LLM-based approaches with program analysis for decomposition~\cite{nitin2025c2saferrust, shetty2024syzygy, shiraishi2024context, nitin2024spectra, evoc2rust, alphatrans} and verification techniques such as formal verification~\cite{yang2024vert}, fuzzing~\cite{eniser2024flourine}, and symbolic testing~\cite{bai2025rustassure}. Despite these advances, these approaches are typically evaluated on relatively small programs (e.g., under 600 LoC), limiting their applicability to realistic codebases.

\textbf{Agentic Approaches for Code Translation.}
In parallel, emerging work explores the use of LLM-powered agents for translation and testing. However, these approaches are either restricted to CLI-based programs~\cite{li2025adversarial}, focus primarily on refactoring C2Rust outputs~\cite{sim2025largelanguagemodelpoweredagent}, are limited to verification of the final translated code~\cite{matchfixagent}, or lack evaluation on diverse datasets~\cite{wangrustify}.

In contrast, our work aims to address these gaps by providing a fully autonomous agentic framework that is not restricted to specific classes of C applications. We exclude multi-threaded and GUI-based programs due to their inherent non-determinism and the additional challenges they introduce for verification. We further evaluate our approach on a rigorous and diverse benchmark.
\section{Conclusion}


We presented \orbit{}, an agentic framework for autonomous C-to-Rust translation that addresses key limitations of prior work, including evaluation on small benchmarks and reliance on brittle static analysis requiring human intervention. \orbit{} decomposes translation into coordinated stages with integrated verification to mitigate agent hallucination. \orbit{} achieves 100\% compilation success and 91.7\% test success, substantially outperforming C2Rust and the CRUST-Bench pipeline in correctness and memory safety. Overall, our results suggest that agentic workflows are a promising direction for applying LLMs to large-scale software modernization.

\section{Data Availability Statement}
 Our source code and dataset are available at \url{https://anonymous.4open.science/r/orbit_c_rust_2026}. The artifacts will be released publicly upon acceptance of the paper.

\bibliographystyle{unsrt}
\bibliography{references}

@misc{immunant2022c2rust,
  author    = {Immunant},
  title     = {{C2Rust}},
  year      = {2022},
  howpublished = {\url{https://github.com/immunant/c2rust}},
  note      = {Accessed: [Insert Date Here]}
}

@inproceedings{crown,
  title={Ownership guided {C} to {R}ust translation},
  author={Zhang, Hanliang and David, Cristina and Yu, Yijun and Wang, Meng},
  booktitle={International Conference on Computer Aided Verification},
  pages={459--482},
  year={2023},
  organization={Springer}
}

@inproceedings{hong2023concrat,
  title={Concrat: An automatic {C-to-Rust} lock {API} translator for concurrent programs},
  author={Hong, Jaemin and Ryu, Sukyoung},
  booktitle={2023 IEEE/ACM 45th International Conference on Software Engineering (ICSE)},
  pages={716--728},
  year={2023},
  organization={IEEE}
}

@article{hong2024don,
  title={Don't Write, but Return: Replacing Output Parameters with Algebraic Data Types in {C-to-Rust} Translation},
  author={Hong, Jaemin and Ryu, Sukyoung},
  journal={Proceedings of the ACM on Programming Languages},
  volume={8},
  number={PLDI},
  pages={716--740},
  year={2024},
  publisher={ACM New York, NY, USA}
}

@inproceedings{pan2024lost,
  title={Lost in translation: A study of bugs introduced by large language models while translating code},
  author={Pan, Rangeet and Ibrahimzada, Ali Reza and Krishna, Rahul and Sankar, Divya and Wassi, Lambert Pouguem and Merler, Michele and Sobolev, Boris and Pavuluri, Raju and Sinha, Saurabh and Jabbarvand, Reyhaneh},
  booktitle={Proceedings of the IEEE/ACM 46th International Conference on Software Engineering},
  pages={1--13},
  year={2024}
}

@article{yang2024unitrans,
  title={Exploring and unleashing the power of large language models in automated code translation},
  author={Yang, Zhen and Liu, Fang and Yu, Zhongxing and Keung, Jacky Wai and Li, Jia and Liu, Shuo and Hong, Yifan and Ma, Xiaoxue and Jin, Zhi and Li, Ge},
  journal={Proceedings of the ACM on Software Engineering},
  volume={1},
  number={FSE},
  pages={1585--1608},
  year={2024},
  publisher={ACM New York, NY, USA}
}

@article{yang2024vert,
  title={VERT: Verified equivalent {Rust} transpilation with large language models as few-shot learners},
  author={Yang, Aidan ZH and Takashima, Yoshiki and Paulsen, Brandon and Dodds, Josiah and Kroening, Daniel},
  journal={arXiv preprint arXiv:2404.18852},
  year={2024}
}

@article{eniser2024flourine,
  title={Towards translating real-world code with {LLMs}: A study of translating to {Rust}},
  author={Eniser, Hasan Ferit and Zhang, Hanliang and David, Cristina and Wang, Meng and Christakis, Maria and Paulsen, Brandon and Dodds, Joey and Kroening, Daniel},
  journal={arXiv preprint arXiv:2405.11514},
  year={2024}
}

@article{nitin2024spectra,
  title={Spectra: Enhancing the code translation ability of language models by generating multi-modal specifications},
  author={Nitin, Vikram and Krishna, Rahul and Ray, Baishakhi},
  journal={arXiv preprint arXiv:2405.18574},
  year={2024}
}

@article{nitin2025c2saferrust,
  title={{C2SaferRust}: Transforming {C} Projects into Safer {Rust} with NeuroSymbolic Techniques},
  author={Nitin, Vikram and Krishna, Rahul and Valle, Luiz Lemos do and Ray, Baishakhi},
  journal={arXiv preprint arXiv:2501.14257},
  year={2025}
}

@article{shetty2024syzygy,
  title={Syzygy: Dual Code-Test {C} to (safe) {Rust} Translation using {LLMs} and Dynamic Analysis},
  author={Shetty, Manish and Jain, Naman and Godbole, Adwait and Seshia, Sanjit A and Sen, Koushik},
  journal={arXiv preprint arXiv:2412.14234},
  year={2024}
}

@article{shiraishi2024context,
  title={Context-aware Code Segmentation for {C-to-Rust} Translation using Large Language Models},
  author={Shiraishi, Momoko and Shinagawa, Takahiro},
  journal={arXiv preprint arXiv:2409.10506},
  year={2024}
}

@article{li2024userstudy,
  title={Translating {C} to {Rust}: Lessons from a user study},
  author={Li, Ruishi and Wang, Bo and Li, Tianyu and Saxena, Prateek and Kundu, Ashish},
  journal={arXiv preprint arXiv:2411.14174},
  year={2024}
}

@misc{tree-sitter,
  title        = {Tree-sitter},
  year         = {2023},
  howpublished = {\url{https://github.com/tree-sitter/tree-sitter}},
  note         = {Accessed: March 14, 2025}
}

@inproceedings{hong2024tag,
  title={To Tag, or Not to Tag: Translating {C}'s Unions to {R}ust's Tagged Unions},
  author={Hong, Jaemin and Ryu, Sukyoung},
  booktitle={Proceedings of the 39th IEEE/ACM International Conference on Automated Software Engineering},
  pages={40--52},
  year={2024}
}

@article{zhang2024scalable,
  title={Scalable, validated code translation of entire projects using large language models},
  author={Zhang, Hanliang and David, Cristina and Wang, Meng and Paulsen, Brandon and Kroening, Daniel},
  journal={arXiv preprint arXiv:2412.08035},
  year={2024}
}

@article{safer-rust,
  author = {Emre, Mehmet and Schroeder, Ryan and Dewey, Kyle and Hardekopf, Ben},
  title = {Translating {C} to Safer {R}ust},
  year = {2021},
  volume = {5},
number = {OOPSLA},
journal = {Proc. ACM Program. Lang.},
month = {oct},
articleno = {121},
}

@INPROCEEDINGS{hongc2rust,
  author={Hong, Jaemin},
  booktitle={Proceedings of the 45th IEEE/ACM International Conference on Software Engineering (ICSE)}, 
  title={Improving Automatic {C}-to-{R}ust Translation with Static Analysis}, 
  year={2023},
  pages={273--277},
}

@article{HUSEIN2025103917,
title = {Large language models for code completion: A systematic literature review},
journal = {Computer Standards \& Interfaces},
volume = {92},
pages = {103917},
year = {2025},
issn = {0920-5489},
doi = {https://doi.org/10.1016/j.csi.2024.103917},
url = {https://www.sciencedirect.com/science/article/pii/S0920548924000862},
author = {Rasha Ahmad Husein and Hala Aburajouh and Cagatay Catal},
}

@article{xu2024large,
  title={Large language models for cyber security: A systematic literature review},
  author={Xu, HanXiang and Wang, ShenAo and Li, Ningke and Wang, Kailong and Zhao, Yanjie and Chen, Kai and Yu, Ting and Liu, Yang and Wang, HaoYu},
  journal={arXiv preprint arXiv:2405.04760},
  year={2024}
}

@inproceedings{bozkir2024embedding,
  title={Embedding large language models into extended reality: Opportunities and challenges for inclusion, engagement, and privacy},
  author={Bozkir, Efe and {\"O}zdel, S{\"u}leyman and Lau, Ka Hei Carrie and Wang, Mengdi and Gao, Hong and Kasneci, Enkelejda},
  booktitle={Proceedings of the 6th ACM Conference on Conversational User Interfaces},
  pages={1--7},
  year={2024}
}

@article{kim2025explainablexrunderstandinguser,
  title={Explainable {XR}: Understanding User Behaviors of {XR} Environments using {LLM}-assisted Analytics Framework}, 
  author={Yoonsang Kim and Zainab Aamir and Mithilesh Singh and Saeed Boorboor and Klaus Mueller and Arie E. Kaufman},
  journal={IEEE Transactions on Visualization and Computer Graphics},
  year={2025}
}

@article{sapkota2025image,
  title={Image, Text, and Speech Data Augmentation using Multimodal {LLMs} for Deep Learning: A Survey},
  author={Sapkota, Ranjan and Raza, Shaina and Shoman, Maged and Paudel, Achyut and Karkee, Manoj},
  journal={arXiv preprint arXiv:2501.18648},
  year={2025}
}

@inproceedings{muzammil2025babylon,
  title={{The Poorest Man in Babylon: A Longitudinal Study of Cryptocurrency Investment Scams}},
  author={Muhammad Muzammil and Abisheka Pitumpe and Xigao Li and Amir Rahmati and Nick Nikiforakis},
  booktitle={Proceedings of The Web Conference (WWW)},
  year={2025}
}

@inproceedings{talha2025lowering,
  title={Lowering Barriers to CAD Adoption: A Comparative Study of Augmented Reality-Based CAD (AR-CAD) and a Traditional CAD Tool},
  author={Talha, Muhammad and Mohiuddin, Abdullah and Javed, Sehrish and Qureshi, Ahmed},
  booktitle={International Design Engineering Technical Conferences and Computers and Information in Engineering Conference},
  volume={89206},
  pages={V02AT02A018},
  year={2025},
  organization={American Society of Mechanical Engineers}
}

@article{talha2026extending,
  title   = {Extending the Cognitive Domain of Bloom’s Taxonomy using Machine Learning},
  author  = {Talha, Muhammad and Shi, Jingchuan and Qureshi, Ahmed},
  journal = {Research Square (Preprint)},
  year    = {2026},
  doi     = {10.21203/rs.3.rs-8704766/v1},
  url     = {https://doi.org/10.21203/rs.3.rs-8704766/v1}
}

@article{jelodar2025large,
  title={Large Language Models ({LLMs}) for Source Code Analysis: applications, models and datasets},
  author={Jelodar, Hamed and Meymani, Mohammad and Razavi-Far, Roozbeh},
  journal={arXiv preprint arXiv:2503.17502},
  year={2025}
}

@inproceedings{hassnain2024counterexamples,
  title={Counterexamples in Safe {Rust}},
  author={Hassnain, Muhammad and Stanford, Caleb},
  booktitle={Proceedings of the 39th IEEE/ACM International Conference on Automated Software Engineering Workshops},
  pages={128--135},
  year={2024}
}

@article{sactor,
  title={LLM-Driven Multi-step Translation from C to Rust using Static Analysis},
  author={Zhou, Tianyang and Lin, Haowen and Jha, Somesh and Christodorescu, Mihai and Levchenko, Kirill and Chandrasekaran, Varun},
  journal={arXiv preprint arXiv:2503.12511},
  year={2025}
}

@article{bai2025rustassure,
  title={RustAssure: Differential Symbolic Testing for LLM-Transpiled C-to-Rust Code},
  author={Bai, Yubo and Palit, Tapti},
  journal={arXiv preprint arXiv:2510.07604},
  year={2025}
}

@misc{evoc2rust,
      title={EvoC2Rust: A Skeleton-guided Framework for Project-Level C-to-Rust Translation}, 
      author={Chaofan Wang and Tingrui Yu and Chen Xie and Jie Wang and Dong Chen and Wenrui Zhang and Yuling Shi and Xiaodong Gu and Beijun Shen},
      year={2025},
      eprint={2508.04295},
      archivePrefix={arXiv},
      primaryClass={cs.SE},
      url={https://arxiv.org/abs/2508.04295}, 
}

@misc{sim2025largelanguagemodelpoweredagent,
      title={Large Language Model-Powered Agent for C to Rust Code Translation}, 
      author={HoHyun Sim and Hyeonjoong Cho and Yeonghyeon Go and Zhoulai Fu and Ali Shokri and Binoy Ravindran},
      year={2025},
      eprint={2505.15858},
      archivePrefix={arXiv},
      primaryClass={cs.PL},
      url={https://arxiv.org/abs/2505.15858}, 
}

@article{alphatrans,
  title={AlphaTrans: A Neuro-Symbolic Compositional Approach for Repository-Level Code Translation and Validation},
  author={Ibrahimzada, Ali Reza and Ke, Kaiyao and Pawagi, Mrigank and Abid, Muhammad Salman and Pan, Rangeet and Sinha, Saurabh and Jabbarvand, Reyhaneh},
  journal={Proceedings of the ACM on Software Engineering},
  volume={2},
  number={FSE},
  pages={2454--2476},
  year={2025},
  publisher={ACM New York, NY, USA}
}

@article{li2025adversarial,
  title={Adversarial Agent Collaboration for C to Rust Translation},
  author={Li, Tianyu and Li, Ruishi and Wang, Bo and Paulsen, Brandon and Mathur, Umang and Saxena, Prateek},
  journal={arXiv preprint arXiv:2510.03879},
  year={2025}
}

@article{matchfixagent,
  title={MatchFixAgent: Language-Agnostic Autonomous Repository-Level Code Translation Validation and Repair},
  author={Ibrahimzada, Ali Reza and Paulsen, Brandon and Jabbarvand, Reyhaneh and Dodds, Joey and Kroening, Daniel},
  journal={arXiv preprint arXiv:2509.16187},
  year={2025}
}

@article{kahn1962topological,
  title={Topological sorting of large networks},
  author={Kahn, Arthur B},
  journal={Communications of the ACM},
  volume={5},
  number={11},
  pages={558--562},
  year={1962},
  publisher={ACM New York, NY, USA}
}

@inproceedings{hong2024metagpt,
  title={{MetaGPT}: Meta Programming for a Multi-Agent Collaborative Framework},
  author={Hong, Sirui and Zhuge, Mingchen and Chen, Jonathan and others},
  booktitle={International Conference on Learning Representations},
  year={2024}
}

@article{huang2023agentcoder,
  title={{AgentCoder}: Multi-Agent-based Code Generation with Iterative Testing and Optimisation},
  author={Huang, Dong and Bu, Qingwen and Zhang, Jie M. and Luck, Michael and Cui, Heming},
  journal={arXiv preprint arXiv:2312.13010},
  year={2023}
}

@inproceedings{qian2024chatdev,
  title={{ChatDev}: Communicative Agents for Software Development},
  author={Qian, Chen and Liu, Wei and Liu, Hongzhang and others},
  booktitle={Proceedings of the Annual Meeting of the Association for Computational Linguistics},
  year={2024}
}

@manual{gcc_nested_functions,
  title        = {{GNU C} Compiler Extensions: Nested Functions},
  author       = {{Free Software Foundation}},
  organization = {Free Software Foundation},
  year         = {2024},
  url          = {https://gcc.gnu.org/onlinedocs/gcc/Nested-Functions.html},
  note         = {Accessed: 2025}
}

@article{khatry2025crust,
  title={Crust-bench: A comprehensive benchmark for c-to-safe-rust transpilation},
  author={Khatry, Anirudh and Zhang, Robert and Pan, Jia and Wang, Ziteng and Chen, Qiaochu and Durrett, Greg and Dillig, Isil},
  journal={arXiv preprint arXiv:2504.15254},
  year={2025}
}

@article{farrukh2025safetrans,
  title={Safetrans: Llm-assisted transpilation from c to rust},
  author={Farrukh, Muhammad and Shah, Smeet and Coskun, Baris and Polychronakis, Michalis},
  journal={arXiv preprint arXiv:2505.10708},
  year={2025}
}

@misc{OpenAIPricing2026,
  author       = {{OpenAI}},
  title        = {API Pricing},
  howpublished = {\url{https://developers.openai.com/api/docs/pricing}},
  year         = {2026},
  note         = {Accessed: 2026-03-18}
}

@misc{AWSBedrockPricing2026,
  author       = {{Amazon Web Services}},
  title        = {Amazon Bedrock Pricing},
  howpublished = {\url{https://aws.amazon.com/bedrock/pricing/}},
  year         = {2026},
  note         = {Accessed: 2026-03-18}
}

@misc{DARPA_TRACTOR,
  author       = {{Defense Advanced Research Projects Agency (DARPA)}},
  title        = {{TRACTOR: Translating All C to Rust}},
  howpublished = {\url{https://www.darpa.mil/research/programs/translating-all-c-to-rust}},
  year         = {2024},
  note         = {Accessed: 2026-03-20}
}

@techreport{TRACTOR_Report_2024,
  author      = {{DARPA TRACTOR Program}},
  title       = {{First TRACTOR Evaluation Report}},
  institution = {Defense Advanced Research Projects Agency (DARPA)},
  year        = {2024},
  type        = {Evaluation Report},
  url         = {https://github.com/DARPA-TRACTOR-Program/Reports/blob/main/First_TRACTOR_Evaluation_Report.pdf},
  note        = {Available via the official TRACTOR Program GitHub repository}
}

@inproceedings{cai2025rustmap,
  title={Rustmap: Towards project-scale c-to-rust migration via program analysis and llm},
  author={Cai, Xuemeng and Liu, Jiakun and Huang, Xiping and Yu, Yijun and Wu, Haitao and Li, Chunmiao and Wang, Bo and Yusuf, Imam Nur Bani and Jiang, Lingxiao},
  booktitle={International Conference on Engineering of Complex Computer Systems},
  pages={283--302},
  year={2025},
  organization={Springer}
}

@misc{OpenAICodex,
  author       = {{OpenAI}},
  title        = {Codex: An {AI} Coding Partner},
  howpublished = {\url{https://openai.com/codex/}},
  year         = {2021},
  note         = {Accessed: 2026-03-23}
}

@misc{OpenCodeAI,
  author       = {{Anomaly Co.}},
  title        = {OpenCode: The Open-Source {AI} Coding Agent},
  howpublished = {\url{https://opencode.ai/}},
  year         = {2025},
  note         = {Accessed: 2026-03-23}
}

@article{shiraishi2024smartc2rust,
  title={SmartC2Rust: Iterative, Feedback-Driven C-to-Rust Translation via Large Language Models for Safety and Equivalence},
  author={Shiraishi, Momoko and Cao, Yinzhi and Shinagawa, Takahiro},
  journal={arXiv preprint arXiv:2409.10506},
  year={2024}
}

@misc{MicrosoftMSRC2019,
  author       = {Gavin Thomas},
  title        = {A proactive approach to more secure code},
  howpublished = {Microsoft Security Response Center (MSRC) Blog},
  year         = {2019},
  month        = {July},
  url          = {https://www.microsoft.com/en-us/msrc/blog/2019/07/a-proactive-approach-to-more-secure-code},
  note         = {Accessed: 2026-03-25}
}

@inproceedings{10.1145/3735544.3735582,
author = {Xu, Qingxiao and Huang, Jeff},
title = {Optimizing Type Migration for LLM-Based C-to-Rust Translation: A Data Flow Graph Approach},
year = {2025},
isbn = {9798400719226},
publisher = {Association for Computing Machinery},
address = {New York, NY, USA},
url = {https://doi.org/10.1145/3735544.3735582},
doi = {10.1145/3735544.3735582},
booktitle = {Proceedings of the 14th ACM SIGPLAN International Workshop on the State Of the Art in Program Analysis},
pages = {8–14}
}

@article{robbes2026agentic,
  title={Agentic Much? Adoption of Coding Agents on GitHub},
  author={Robbes, Romain and Matricon, Th{\'e}o and Degueule, Thomas and Hora, Andre and Zacchiroli, Stefano},
  journal={arXiv preprint arXiv:2601.18341},
  year={2026}
}

@article{li2025rise,
  title={The rise of ai teammates in software engineering (se) 3.0: How autonomous coding agents are reshaping software engineering},
  author={Li, Hao and Zhang, Haoxiang and Hassan, Ahmed E},
  journal={arXiv preprint arXiv:2507.15003},
  year={2025}
}

@article{BusinessInsiderUberAI2026,
  author       = {Business Insider Staff},
  title        = {{AI} Coding Boom Shifts Software Developers Toward Management},
  journal      = {Business Insider},
  year         = {2026},
  month        = {March},
  day          = {18},
  url          = {https://www.businessinsider.com/ai-coding-changing-software-developer-role-2026-3},
  note         = {Accessed: 2026-03-25}
}

@article{gao2025pr2,
  title={Pr2: Peephole raw pointer rewriting with llms for translating c to safer rust},
  author={Gao, Yifei and Wang, Chengpeng and Huang, Pengxiang and Liu, Xuwei and Zheng, Mingwei and Zhang, Xiangyu},
  journal={arXiv preprint arXiv:2505.04852},
  year={2025}
}

@article{hong2025type,
  title={Type-migrating C-to-Rust translation using a large language model},
  author={Hong, Jaemin and Ryu, Sukyoung},
  journal={Empirical Software Engineering},
  volume={30},
  number={1},
  pages={3},
  year={2025},
  publisher={Springer}
}

@misc{TheGreatRefactor2024,
  author       = {TRACTOR Program Developers},
  title        = {The Great Refactor: {DARPA TRACTOR} Documentation and Resources},
  howpublished = {\url{https://www.thegreatrefactor.org/}},
  year         = {2024},
  note         = {Accessed: 2026-03-25}
}

@misc{hassnain2026cargosherlocksmtbasedchecker,
      title={Cargo Sherlock: An SMT-Based Checker for Software Trust Costs}, 
      author={Muhammad Hassnain and Anirudh Basu and Ethan Ng and Caleb Stanford},
      year={2026},
      eprint={2512.12553},
      archivePrefix={arXiv},
      primaryClass={cs.LO},
      url={https://arxiv.org/abs/2512.12553}, 
}

@inproceedings{babar2025open,
  title={Open-Source LLMs for Technical Q\&A: Lessons from StackExchange},
  author={Babar, Zeerak and Khan, Nafiz Imtiaz and Hassnain, Muhammad and Filkov, Vladimir},
  booktitle={International Conference on Software Engineering of Emerging Technology},
  pages={615--626},
  year={2025},
  organization={Springer}
}

@misc{AWS_Kani_2022,
  author       = {Zeljic, Aleksandar and Taneja, Shaobo and Tomb, Aaron},
  title        = {Verify the safety of the {Rust} standard library},
  howpublished = {AWS Open Source Blog},
  year         = {2022},
  month        = {July},
  url          = {https://aws.amazon.com/blogs/opensource/verify-the-safety-of-the-rust-standard-library/},
  note         = {Accessed: 2026-03-25}
}

@article{wangrustify,
  title={Rustify: Towards Repository-Level C to Safer Rust via Workflow-Guided Multi-Agent Transpiler},
  author={Wang, Chen and Huang, Yujun and Li, Peng and Gong, Lina and Wu, Fei}
}

\end{document}